\documentclass[aps,pra,twocolumn,showpacs,groupedaddress]{revtex4-1}

\usepackage[latin1]{inputenc}
\usepackage{amsfonts}
\usepackage{amsmath}
\usepackage{graphicx,graphics}

\newcommand{\atanh}{\ensuremath{\mathop{\mathrm{atanh}}}}     
\newcommand{\ket}[1]{\ensuremath{\left|#1\right\rangle}}             
\newcommand{\bra}[1]{\ensuremath{\left\langle#1\right|}}             
\newcommand{\dif}{\ensuremath{{\rm d}}}       
\newcommand{\pr}{\ensuremath{^{\prime}}}
\newcommand{\ketbra}[2]{\ensuremath{\ket{#1\vphantom{#2}}\bra{#2\vphantom{#1}}}} 

\newcommand{\etal}{\emph{et al.}}

\usepackage{color}

\begin{document}

\title{Limited path entanglement percolation in quantum complex networks}
\author{Mart\'\i\ Cuquet}
\author{John Calsamiglia}
\affiliation{Grup de F\'{\i}sica Te\`orica: Informaci\'o i Fen\`omens
Qu\`antics, Universitat Aut\`onoma de Barcelona, 08193 Bellaterra, Barcelona,
Spain}
\date{\today}

\begin{abstract}
We study entanglement distribution in quantum complex networks where nodes are connected by bipartite entangled states. These networks are characterized by a complex structure, which dramatically affects how information is transmitted through them. For pure quantum state links, quantum networks exhibit a remarkable feature absent in classical networks: it is possible to effectively rewire the network by performing local operations on the nodes. We propose a family of such quantum operations that decrease the entanglement percolation threshold of the network and increase the size of the giant connected component. We provide analytic results for complex networks with arbitrary (uncorrelated) degree distribution. These results are in good agreement with numerical simulations, which also show enhancement in correlated and real world networks. The proposed quantum preprocessing strategies are not robust in the presence of noise. However, even when the links consist of (noisy) mixed state links, one can send quantum information through a connecting path with a fidelity that decreases with the path length. In this noisy scenario, complex networks offer a clear advantage over regular lattices, namely the fact that two arbitrary nodes can be connected through a relatively small number of steps, known as the small world effect. We calculate the probability that two arbitrary nodes in the network can successfully communicate with a fidelity above a given threshold. This amounts to working out the classical problem of percolation with limited path length. We find that this probability can be significant even for paths limited to few connections, and that the results for standard (unlimited) percolation are soon recovered if the path length exceeds by a finite amount the average path length, which in complex networks generally scales logarithmically with the size of the network.
\end{abstract}


\maketitle

\section{Introduction}

Networks permeate all informational structures. They underlie natural, social
and artificial systems where different parties interact, describing the flow
of information between them. Differences in the characteristics of such
interactions and how they evolve give growth to different types of structures:
regular lattices, completely random networks or, spanning the range between
these two, complex networks, which do not have a regular structure but neither
are completely random. Quantum information is not an exception, and quantum
networks \cite{Kimble2008} where nodes communicate between them through
quantum channels are essential to quantum information processing and
distributed applications. One of the key tasks in these networks is the
transmission of quantum information between two distant nodes of the network.
This task depends not only on the quality of the connections between nodes and
on the amount of resources, but also on the underlying structure of the
network. Therefore, understanding how structural properties affect the
functionality of the network will allow both the design of better network
architectures and the modification of existing ones that make feasible
communication at further distances, among a greater number of nodes or in the
presence of higher levels of noise.

\begin{figure}[t]
  \begin{center}
    \includegraphics[width=.4\textwidth]{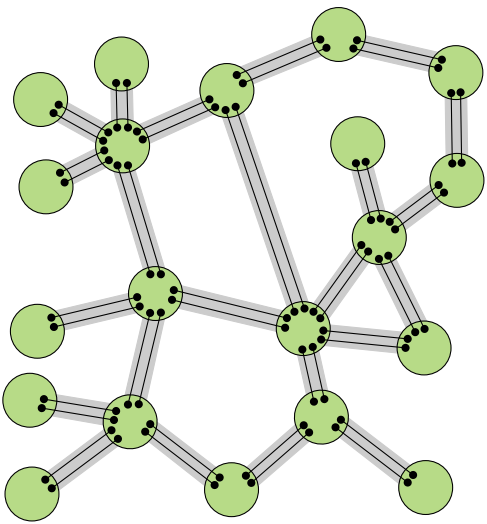}
  \end{center}
  \caption{(Color online) A quantum network. Nodes are big (green) circles.
  Links between nodes are represented in grey, each holding a number of
  bipartite entangled states (black lines) shared between nodes.}
  \label{fig:qnet}
\end{figure}

Two distant nodes in a network may be connected by one path of entangled
states (Figure \ref{fig:qnet}). In this case, long-distance entanglement
between two nodes can be established with a probability that decays
exponentially with the distance separating the nodes. This problem can
be overcome by quantum repeaters, which create a distant entangled pair
of high fidelity \cite{Briegel1998}. However, such technique require a
number of qubits in each node that scales logarithmically with the distance
\cite{Dur1999}. More important, though, is that it only considers a
one-dimensional connection between the two nodes. These two nodes, however,
may be embedded in a more realistic, higher dimensional network. In this
case, a higher number of paths may exist which can help in the communication:
with the existence of clusters of nodes connected by entangled states, two
distant nodes will be able to establish entanglement between them if they both
belong to the same cluster \cite{Calsamiglia2005}. Entanglement percolation,
which makes use of such higher dimensional networks, was first proposed in
the honeycomb lattice \cite{Acin2007} and later extended to other regular
lattices \cite{Perseguers2008, LapeyreJr.2009}, to schemes using multipartite
entanglement \cite{Perseguers2010} and to noisy networks \cite{Broadfoot2009,
Broadfoot2010, Perseguers2010a, Broadfoot2010a}. In Ref.\ \cite{Cuquet2009},
we studied entanglement distribution in a wide class of complex networks
with pure state connections. Complex networks arise in many real scenarios,
notably including the most important real world communication networks, and
it is very plausible that they will become relevant in quantum communication
architectures too. They offer very rich properties and phenomena. Interesting
quantities can be computed requiring only statistical properties. This might
seem a limitation but it can represent an advantage: it makes mathematically
tractable some problems that are hard or impossible to solve on lattices, and
provides a minimal description in scenarios were complete knowledge of the
system is not available or is hard to obtain.

One of the primary features of networks is the presence, or absence, of a
cluster of nodes connected between them (a \emph{connected component}, in
graph theory language) whose size is of the order of the size of the network.
Such cluster is called the giant connected component, and in the asymptotic
limit of infinite size networks it is defined as the cluster spanning a finite
fraction $S$ of nodes of the network. This concept is very closely related to
that of a percolating cluster. In bond percolation, for example, edges are
occupied with some probability $\phi_1$, thus connecting their end vertices, and
empty with probability $1-\phi_1$, disconnecting them. Then there exist a
percolation threshold $\phi_1^*$ in this occupation probability: below the
threshold, all components are of finite size, while above it there exist one
giant connected component whose size is comparable to the network size. This
threshold is the critical point of a phase transition, generally of second
order, and can be manifested by the divergence of the average component size,
which acts like a susceptibility in a magnetic material.

The percolation threshold and the size of the giant connected component, as
well as many other properties, strongly depend on the basic structure of
the network \cite{Moore2000a, Newman2001, Dorogovtsev2008} as well as on
degree-degree correlations \cite{Boguna2003, Goltsev2008} and clustering
\cite{Newman2009}. Therefore, a change in the structure of a network can
affect its ability to communicate information. For example, the scale free
topology of Internet makes it strong---resilient---against the failure of
random nodes \cite{Cohen2000}, but not against target attacks directed to
its major hubs (nodes with highest number of neighbors) \cite{Albert2000,
Cohen2001}. This relation between the structure of the network and the
communication over it can be also exploited to benefit the earlier appearance
of the giant cluster and to find architectures that allow communication even
in the presence of noise.

In this paper we study the distribution of quantum information over quantum
complex networks. We first focus on networks where nodes are connected by
bipartite pure entangled states. We propose a transformation of the network
that, using only local knowledge, can change the structure of the network and
decrease its percolation threshold. We also calculate how the percolation
threshold and the size of the giant component change after the transformation.
Then, we turn to mixed state connections between nodes and show that the small
world behavior of many complex networks allows quantum communication above
some fidelity bound for finite, but very large, quantum complex networks.

\section{\label{sec:rnd_graphs}Random graphs}

A network is naturally represented by a graph $G$, which is an ordered pair
of sets $G = \{V, E\}$. $V$ is the set of vertices (or nodes, or points), and
$E$ the set of edges (or links, or lines), which are pairs of elements of $V$
and represent the connections between them. In this paper we consider always
undirected graphs with neither multiple edges (i.e., either zero or one edge
between every pair of vertices) nor self-loops. The degree of a vertex, $k$,
is the number of edges emerging from it. A connected component, or cluster, is
a subgraph where any two vertices are connected by at least one path of edges
and to which no more vertices can be added without losing this property.

Random graphs \cite{Bollobas2001} are ensembles of graphs $\mathcal{G}$ of
the same size, with a probability $P(G)$ assigned to every graph $G$ in the
ensemble. For any property $O(G)$ of a graph we can calculate its average over
the ensemble:
\begin{equation}
  \langle O \rangle_G = \sum_{G\in\mathcal{G}} O(G)P(G) .
  \label{eq:property_average}
\end{equation}
However, in most real scenarios, but also in many theoretical models, only
a single, large graph is studied. Such graphs are said to be self-averaging
if the property we are studying is well characterized by its mean. This
happens when the graph is large enough to make fluctuations around the average
vanish. For a more detailed discussion about self-averaging, see e.g.\ Ref.\
\cite{Bialas2008} for random graphs and \cite{Serrano2007} for the World Wide
Web network. We will consider this assumption in the following, and check its
validity numerically in the examples we consider.

One of the basic properties of a graph is the distribution of the probability
$p_k$ that a vertex has degree $k$. There is also a related distribution
that will come in handy later, and is that of the excess degree: the number
of edges, $k$, emerging from a vertex reached through another edge, and
excluding it. This probability can be found easily by considering first the
degree of a vertex reached through an edge. Since vertices with higher degree
are easier to reach, such probability is proportional to the degree of the
reached vertex, $kp_k/\langle k\rangle$. The excess degree probability $r_k$
is therefore
\begin{equation}
  r_k = \frac{(k+1)p_{k+1}}{\langle k \rangle}.
  \label{eq:excess_degree}
\end{equation}

Generating functions \cite{Wilf2006} are a mathematical tool that shows very
useful when studying properties of graphs described by probability
distributions \cite{Newman2001}. Among other useful properties, they allow for
a straightforward convolution of distributions. Let us introduce them using
the degree distribution. The function $g_p(x)$ that generates the distribution
$\{p_k\}$ is the power series of $x$ with coefficients equal to the
probabilities in the distribution:
\begin{equation}
  g_p(x) = \sum_{k\ge0} p_k x^k.
  \label{eq:degree_gf}
\end{equation}
Note that each probability $p_k$ can be recovered from its generating function
(\ref{eq:degree_gf}) by taking the $k$-th derivative of $g_p(x)$ at $x=0$,
\begin{equation}
  p_k = \frac{1}{k!} \left. \frac{ \dif^k g_p(x)}{ \dif x^k} \right|_{x=0}.
  \label{eq:gf_recover_prob}
\end{equation}
Since the probability distribution is normalized, $\sum_k p_k = 1$, so is its
generating function, $g_p(1) = 1$. It is also convergent for $|x|\le1$, which
is all what we will use here.

The first moment of the distribution $p_k$, which corresponds to the average
degree of the graph $\langle k \rangle$, is equal to the first derivative at
$x=1$:
\begin{equation}
  \langle k \rangle = g_p\pr(1) = \sum_{k\ge1} kp_k.
  \label{eq:gf_first_moment}
\end{equation}
This allows to express the generating function for the excess degree
distribution, $g_r(x)$, in terms of (\ref{eq:degree_gf}):
\begin{equation}
  g_r(x) = \sum_{k\ge0} r_kx^k = \frac{g_p\pr(x)}{g_p\pr(1)}.
  \label{eq:excess_degree_gf}
\end{equation}
Higher moments can be similarly found by taking more derivatives. In general,
the $n$-th moment is
\begin{equation}
  \langle k^n \rangle = \sum_{k\ge0} k^np_k = \left[ \left(
  x\frac{\dif}{ \dif x} \right)^n g_p(x) \right]_{x=1}.
  \label{gf_nth_moment}
\end{equation}

Convolution of independent distributions can be obtained by multiplication of
their respective generating functions. For example, the total number of edges
emerging from $n$ independent vertices (the sum of their degrees) is generated
by $\left[ g_p(x) \right]^n$.

\section{\label{sec:net_examples}Network examples}

In this paper we calculate properties such as the average component size,
the giant connected component size and the percolation threshold. Analytical
results are found for random networks with uncorrelated degree distribution,
and for the Watts--Strogatz small world model in the mixed state scenario. We
also discuss several network topologies as concrete examples of our results:
the simple Bethe lattice, two networks (Erd\H os--R\'enyi and scale free)
belonging to the configuration model, the Watts--Strogatz small world model
and two real world networks. All of them share a common property known as the
``small world effect'': the average path length, or intervertex distance,
scales logarithmically with the size of the network, rather than as a positive
power of the size, $N^{1/d}$, as is the case in finite-dimensional networks.
Here we present a short description of each of these network models.

\begin{figure}[t]
  \begin{center}
    \includegraphics[width=.22\textwidth]{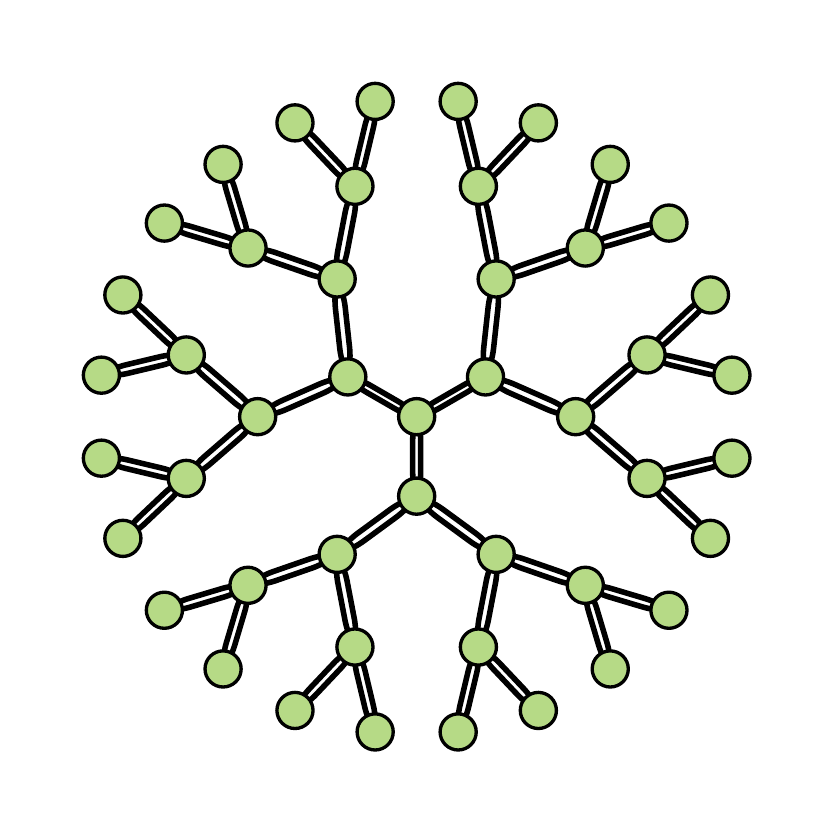}
    \includegraphics[width=.22\textwidth]{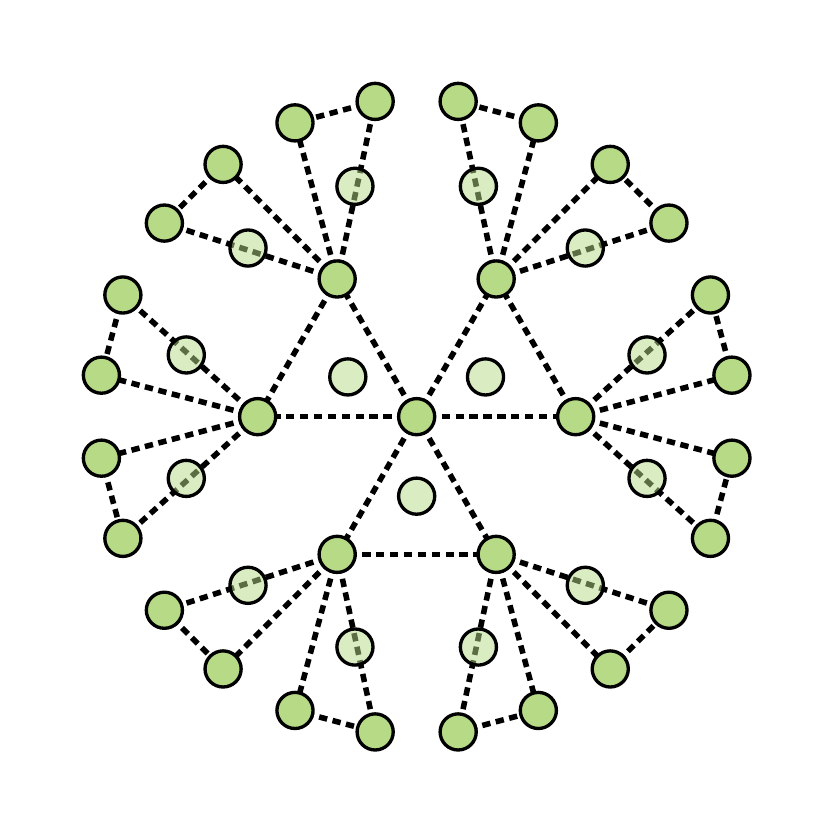}
  \end{center}
  \caption{(Color online) Example of the first five crowns of a Bethe lattice
  with coordination number $k=3$. Left: Bethe lattice before $q$-swap. Right:
  Same network after applying 3-swap.}
  \label{fig:bethe_lattice}
\end{figure}

Let us start with the simplest network example. The \emph{Bethe lattice} is
not a random graph but has some similar properties such as a local tree-like
structure and the small world effect (as long as the degree of its vertices
exceeds 2), while at the same time remains amenable to analytical study.
A Bethe lattice with coordination number $k$ is defined as an infinite
regular graph where every vertex has the same degree $k$ and is topologically
equivalent to all the others, as shown in Figure~\ref{fig:bethe_lattice}.
Random regular graphs---graphs where all vertices have a fixed degree but
edges are placed randomly---asymptotically approach Bethe lattices, making
them a relevant model where analytical treatment is usually possible.

\begin{figure}[t]
  \begin{center}
    \includegraphics[width=.22\textwidth]{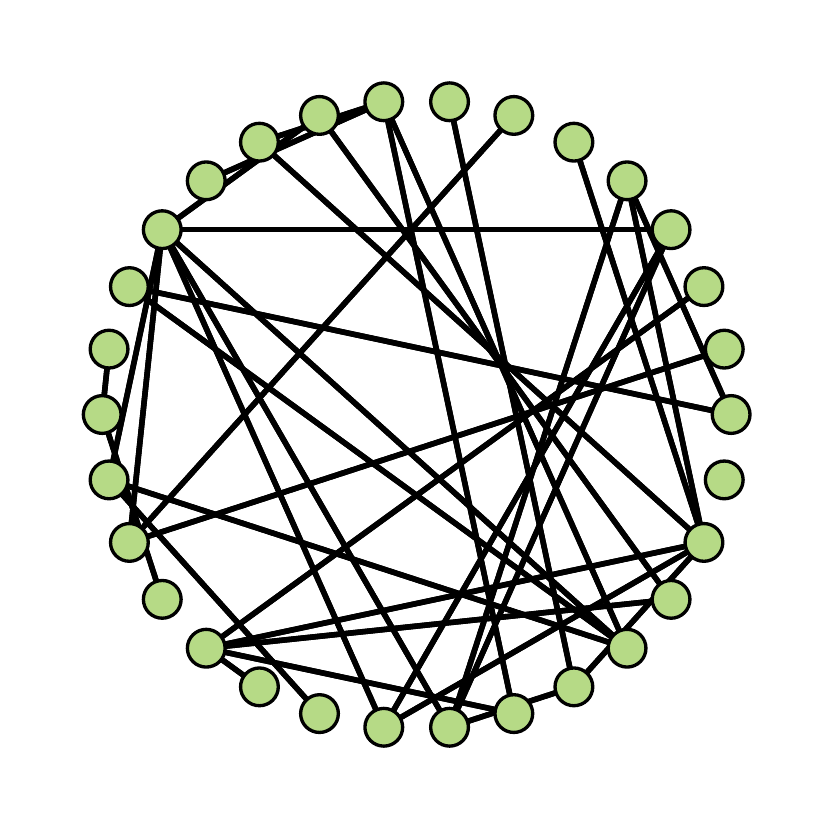}
    \includegraphics[width=.22\textwidth]{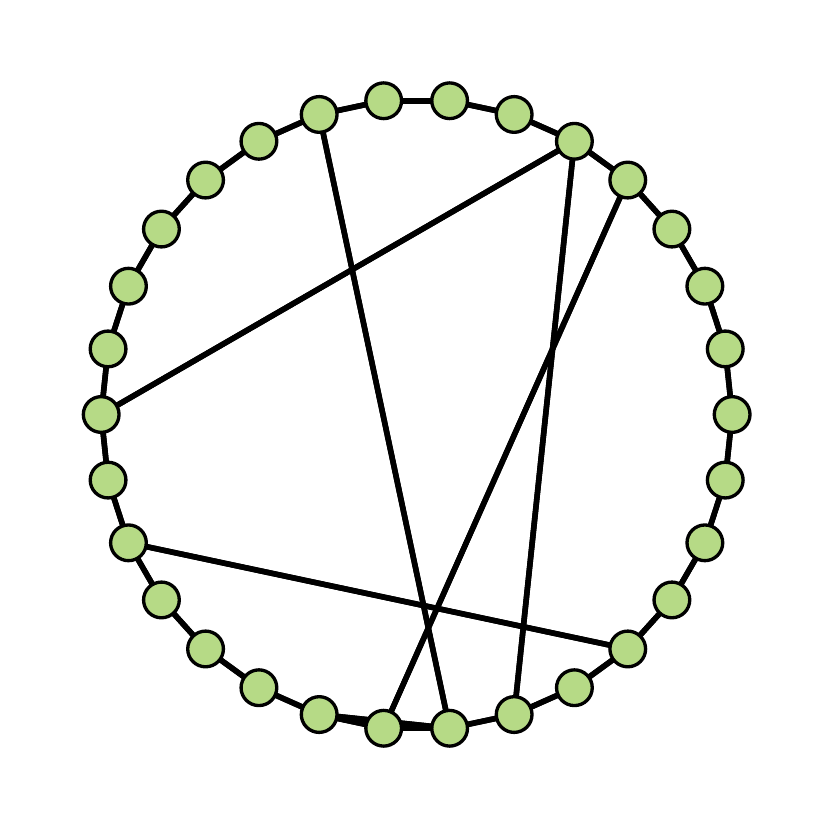}
  \end{center}
  \caption{(Color online) Left: Example of an Erd\H os--R\'enyi network with
  $N=30$ and $z=4/3$. Right: Example of a Watts--Strogatz network with $N=30$
  and $\beta=0.2$.}
  \label{fig:er_swas_network}
\end{figure}

\emph{Erd\H os--R\'enyi} graphs \cite{Gilbert1959, Erdos1959, Erdos1960}
are maximally random graphs with the only constrain $\langle k \rangle
= z$. An Erd\H os--R\'enyi network with $N$ vertices can be realized by
randomly placing $M=Nz/2$ edges, or similarly by placing an edge between every
pair of vertices with probability $z/N$ (also known as the Gilbert model),
which is asymptotically equivalent \cite{Dorogovtsev2008}.
Figure~\ref{fig:er_swas_network} shows an example of a small Erd\H os--R\'enyi
network. Their degree distribution is Poissonian, $p_k=e^{-z}z^k/k!$, with
generating functions $g_p(x)=g_r(x)=\exp[z(x-1)]$.

Real world networks are not Poissonian but typically exhibit a power-law
(scale free) degree distribution, $p_k\sim k^{-\tau}$, characterized by a
relatively important number of nodes with a degree much greater than the
average. Scale free networks with $\tau\le3$ have a percolation threshold
at $\phi_1=0$, while for networks with $\tau>3$ a finite threshold appears.
However, in heavy-tailed networks such like these, a cutoff in the degree
naturally appears in scenarios where high degrees cannot exist due to, e.g.,
targeted attacks, physical constraints, saturation effects or finite size
networks. For this reason we consider scale free networks with an exponential
cutoff, $p_k=Ck^{-\tau}e^{-k/\kappa}$ ($C$ is a normalizing constant),
while the pure scale free behavior can still be recovered by taking the
limit $\kappa\to\infty$. The cutoff $\kappa$ strongly affects the network
properties, and in particular networks with $\tau\le3$ have now a finite
threshold.

Random graphs with uncorrelated degree distribution such as the previous
models exhibit a very low level of clustering (also known as transitivity):
the likelihood that two neighbors of the same vertex are also neighbors
between them. Aside from regular lattices, which have a high level of
clustering, there are also many real world networks with this property.
This is specially true for social networks, but also for communication and
biological networks. To study this behavior, many models have been proposed,
maybe the most studied being the \emph{Watts--Strogatz model} \cite{Watts1998}.
This model is a random graph with ordered local structure and high level
of clustering but still with surprisingly low average path length (see
Figure \ref{fig:er_swas_network}. Here we study a slight modification, also
considered in \cite{Moore2000a}. It is generated by placing $N$ vertices in a
1D ring. Then, $N$ additional random edges, called ``shortcuts'', are added
with probability $\beta$, thus giving an average of $\beta N$ shortcuts.

We also study two real world networks. The first case represents a real world
scale free network consisting of \emph{World Wide Web} sites in the nd.edu
domain \cite{Albert1999}. In this case we introduce an artificial cutoff by
neglecting nodes with degree $k\ge15$, leaving a graph with 142~192 nodes
and 170~352 edges. The second real world example is the \emph{OpenPGP Web
of Trust}, a social network representing the trust between OpenPGP users.
Without going into much detail, OpenPGP is a standard encryption protocol
for securing email communications using public key cryptography. If Alice
wants to send a secure message to Bob, she has to use Bob's public key to
encrypt it. The authentication problem arises when Alice cannot verify if
the key she is using is really owned by Bob. A solution to this problem is
the Web of Trust, in which every user signs a public key if she trusts it,
thus generating a directed graph. To trust a key, usually a user has to meet
with the key owner and check that he is really who he claims to be. This
social model is thus relevant to quantum communication in the sense that
at this point the two users could create a bipartite entangled state and
then separate, each keeping one of the parts. By repeatedly doing so between
different pairs of users, as in the Web of Trust, a quantum network would be
created. Here we use the strongly connected component of the Web of Trust
obtained from the Swiss keyserver \footnote{wwwkeys.ch.pgp.net:11371/pks/,
public data available at www.lysator.liu.se/\~\ jc/wotsap/index.html.} as of
May 25, 2010, containing 41~459 keys and 424~577 signatures. We considered
only bidirectional edges, corresponding to users who mutually signed their
keys. This leaves an undirected graph with 38~550 keys and 145~388 two-way
signatures.

\section{Pure state networks}

We first focus on networks of pure, nonmaximally entangled states, as in
\cite{Acin2007}. In this case, edges have some probability $\phi_n$ of
being converted into maximally entangled states depending on the amount
of entanglement and the number $n$ of bipartite states per edge. Singlets
can then be used for perfect teleportation, i.e., they are equivalent to a
single-use ideal quantum channel. This strategy can be directly mapped into
a bond percolation problem, and is thus called \emph{classical entanglement
percolation}. There exist then a critical probability $\phi_1^*$, called the
percolation threshold, above which the giant component appears with fractional
size $S>0$. Above this threshold, any two nodes are able to share maximal
entanglement if they both belong to the giant component. This happens with
a probability $S^2$ which is independent of the distance, but that strongly
depends on the network topology. Hence, for a given type of edges $\phi_n$
long-distance entanglement will only be possible for networks fulfilling
$\phi_n>\phi_1^*$. Remarkably, due to the quantum nature of the connections it
is possible to drastically change the network topology by local actions: a
particular measurement is done on qubits within the same node, establishing
new connections between neighboring nodes. Thus, a quantum preprocessing of
the network can be carried before edges are converted into singlets, so the
new structure provides, e.g., a better percolation threshold. Moreover, in
order to carry the particular preprocessing strategies we consider it is not
necessary to know the precise structure of the network. Given only general
statistical properties of the network, we propose strategies that act on each
node depending only on locally accessible information, such as the degree
of the node. We calculate the new percolation threshold $\tilde{\phi}_1^*$
of the modified network and the evolution of the giant connected component
$\tilde{S}$. Above the threshold any two nodes will be able to establish an
entangled state with probability $\tilde{S}^2$, again independently of the
distance between them. Thus, quantum preprocessing can benefit communication
in two ways: by lowering the percolation threshold and by an increase in the
giant component size.

\subsection{Network model}

We consider a quantum network in which neighboring nodes share $n=2$ copies of
a bipartite pure entangled state of two qubits,
\begin{equation}
  \ket{\psi} = \sqrt{\lambda_0}\ket{00} + \sqrt{\lambda_1}\ket{11},
  \label{eq:pure_edge}
\end{equation}
where $\sqrt{\lambda_0}\ge\sqrt{\lambda_1}\ge0$ are its Schmidt coefficients.
A partially entangled state can be converted into a maximally
entangled state (\emph{singlet} for short) with singlet conversion probability
(SCP) that only depends on its largest Schmidt coefficient \cite{Vidal1999}. For the
state $\ket{\psi}$, the SCP is
\begin{equation}
  \phi_1=\min[1,2(1-\lambda_0)].
\end{equation}
We will consider edges that are of the form
$\ket{\psi}^{\otimes 2}$, and thus can be converted to singlets with SCP
\begin{equation}
  \phi_2=\min[1,2(1-\lambda_0^2)].
\end{equation}
With this probability two neighbors can establish a perfect channel between
them. As we discussed above, for two distant nodes this probability depends on
the structure of the network that connects them.

\subsection{Modifying the network: $q$-swap}

\begin{figure}[t]
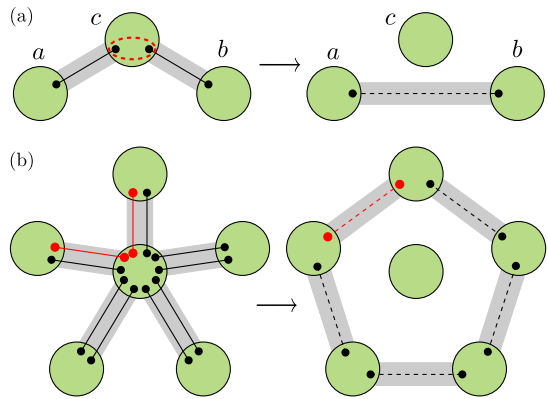

  \begin{center}
    \includegraphics[width=.4\textwidth]{entanglement-swapping-horizontal}
    \includegraphics[width=.4\textwidth]{q-swap-horizontal}
  \end{center}
  \caption{(Color online) Entanglement swapping (a) and $q$-swap (b). Black
  dots represent qubits, big circles nodes, solid lines states $\ket{\psi}$
  and dashed lines the resulting mixed state after entanglement swapping, with
  same SCP as $\ket{\psi}$.}
  \label{fig:q-swap}
\end{figure}

In \cite{Cuquet2009} we introduced a network transformation, the $q$-swap,
that requires only local information of the network: the degree of a target
node and the status of its neighbors. The $q$-swap is built upon a basic
transformation: entanglement swapping or swap \cite{Zukowski1993}. In a
subgraph with three nodes, the party at the target node $c$ performs a Bell
measurement on two qubits, each of them belonging to states $\ket{\psi}$
shared with different nodes $a$ and $b$ (see Figure~\ref{fig:q-swap}). After
this operation, the central qubits become disentangled from $a$ and $b$,
but in return a mixed entangled state with the same SCP as $\ket{\psi}$
is created between $a$ and $b$ \cite{Acin2007}. Note that this operation
can not be repeated with a fourth node because the newborn state shared
between $a$ and $b$ is not of the form of $\ket{\psi}$. The $q$-swap
performs swap transformations between successive pairs of neighbors of
a central target node of degree $q$, thus changing and initial $q$-star
with edges $\ket{\psi}^{\otimes 2}$ to a $q$-cycle with newborn edges,
while the central target node becomes disconnected from the network (see
Figure~\ref{fig:q-swap}). For a given network topology, we will see that
performing $q$-swaps on nodes with certain degrees improves the threshold. It
is worth noting, however, that in some instances the application of particular
$q$-swaps may be counterproductive.

\subsection{Percolation threshold and giant component}

The main two figures of merit that we will use to compare the two strategies
are the percolation thresholds ($\phi_1^*$ and $\tilde{\phi}_1^*$ for the
classic and $q$-swap strategies respectively) and the size of the giant
components ($S$ and $\tilde{S}$). Since $q$-swaps disconnect vertices,
which can be chosen not to be the two corresponding to the parties that
want to communicate, the probability to connect two remote nodes is in fact
$\hat{S}^2$, where $\hat{S}=\tilde{S}S_1/\tilde{S}_1$ and $S_1$ is the value
of $S$ at $\phi_1=1$. The percolation threshold tells us which is the minimum
amount of entanglement needed for long-distance communication, while the
square of the giant connected component size is the probability that any two
nodes can communicate.

To compute these two values we will use the generating function formalism
described in Section~\ref{sec:rnd_graphs}. The key probability distributions
are those of finding a connected component of \emph{finite} size $s$,
either when a random vertex is chosen, $P_s$, or when a random edge is
followed to one of its ends, $R_s$. A random edge is empty with probability
$R_0=1-\phi_2$, giving a cluster of size 0. When it is occupied, then a node
of degree $k+1$ is reached with probability $r_k$, giving access to $k$
clusters. Therefore,
\begin{equation}
  R_{s\ge1} =
  \phi_2 \sum_{k=0}^\infty r_k \sum_{s_1,s_2,\cdots,s_k} R_{s_1} R_{s_2} \cdots
  R_{s_k} \delta_{s,1+\sum_{i=1}^ks_k}.
  \label{eq:component_size_edge}
\end{equation}
We have assumed that components are treelike, i.e., that they do not have
finite loops. This is indeed true for finite components, since an edge exiting
such a component will reconnect back to itself with probability proportional
to $s/N\to0$. The function generating $R_s$ is then $h_R(x) = \sum_{s\ge0}
R_sx^s$, which gives the recurrence relation
\begin{equation}
  h_R(x)
  = 1-\phi_2 + \phi_2 x g_r [h_R(x)].
  \label{eq:component_size_edge_gf}
\end{equation}
It is crucial to notice here that by restricting to finite $s$ we have
explicitly excluded the infinite giant component from $h_R(x)$. Thanks to this
the previous treelike assumption holds.

We can proceed similarly with $P_s$. A random vertex has degree $k$ with
probability $p_k$, giving acces to $k$ clusters. The probability that this
random vertex is in a component of size $s$ is then
\begin{equation}
  P_s = \sum_{s=1}^\infty p_k \sum_{s_1,s_2,\dots,s_k} R_{s_1}R_{s_2}\dots
  R_{s_k} \delta_{s,1+\sum_{i=1}^k s_k}.
  \label{eq:component_size_vertex}
\end{equation}
Now the generating function for $P_s$ is related to $h_R(x)$,
\begin{equation}
  h_P(x)
  = x g_p[h_R(x)].
  \label{eq:component_size_vertex_gf}
\end{equation}

Knowledge of $h_P(x)$ and $h_R(x)$ allows for the derivation of
$\phi_1^*$ and $S$. The probability $u$ that an edge connects to a
finite component is the smallest real solution of $u\equiv h_R(1)$ in
Eq.~(\ref{eq:component_size_edge_gf}), which is in general a transcendental
function. In fact, the percolation threshold is the value of $\phi_1$ at which
a solution $u<1$ appears. Moreover, $u^2$ is the probability that the edge
connects to a finite components through both ends, so with probability $1-u^2$
a random edge belongs to the giant component (which is 0 below the threshold).

Similarly, the main quantity of interest, the giant component size, can
be computed as the missing component in the whole network. The sum of all
$P_s$, $h_P(1)$, gives the probability that a random vertex belongs to a
\emph{finite} component, while the probability $S$ that it is in the giant
component is $S = 1-\sum_{s\ge1}P_s = 1-h_P(1)$. Again, this equation is
usually transcendental and has to be solved numerically. For instance, the
Erd\H os--R\'enyi model, with $g_p(x) = g_r(x) = e^{z(x-1)}$, has a giant
component fraction $S=1-e^{-z\phi_2 S}$ \cite{Newman2001}. In this case the
solution can be expressed in terms of the Lambert $W$ function,
\begin{equation}
  S = 1 + \frac{1}{z\phi_2} W(-z\phi_2 e^{-z\phi_2}),
  \label{eq:erS}
\end{equation}
and the phase transition to $S>0$ occurs at the well-known point $\phi_1^*=1/z$.
On the other hand, first moments can usually be computed even when a
closed expression for $h_P(x)$ and $h_R(x)$ is not known. As a relevant
example, the average component size $\langle s \rangle = h_P\pr(1)$ is an
important property of the network that provides an alternative way of finding
the probability threshold: it is at this point that $\langle s \rangle$
diverges. From the derivatives of Eqs.~(\ref{eq:component_size_edge_gf})
and (\ref{eq:component_size_vertex_gf}) it is immediate to find that this
divergence can be traced back to that of
\begin{equation}
  h_R\pr(1)=\frac{\phi_2}{1-\phi_2 g_r\pr(1)}.
\end{equation}
This brings the general result for the critical SCP $\phi_2^{*} = 1/g_r\pr(1)$
\cite{Callaway2000}.

\begin{figure}[t]
  \begin{center}
    \includegraphics[width=.4\textwidth]{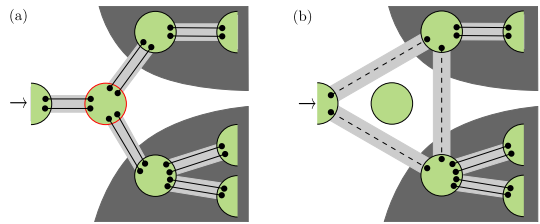}
  \end{center}
  \caption{(Color online) Example of the branching process before and after
  a $3$-swap, starting at the leftmost node. (a) Before any operation the
  branching process arrives at a node of degree 3, leading to 2 components
  (in dark grey). (b) After the 3-swap, the branching process is already in a
  3-cycle, each of its nodes belonging to one of the 2 components.}
  \label{fig:branching_qswap}
\end{figure}

We now want to understand how the $q$-swap transformation changes the
percolation properties of the network. Every particular $q$-swap can be
implemented (or not) with probability $\Pi_q$ (or $1-\Pi_q$) on nodes
of degree $q$. Giving the values for each $\Pi_q$ specifies the quantum
strategy. $q$-swaps introduce cycles, so components are no longer treelike
and generating functions can not be directly used. Note however that, since
newborn edges cannot be reused, those cycles do not overlap between each other
and can thus be treated as blocks of a treelike component by considering two
steps in the branching process. We first compute the generating function for
the probability $R_s$ \emph{after} $q$-swaps are done, $\tilde{h}_R(x)$.
Now, instead of arriving to a vertex of degree $q$ connecting to other $q-1$
components, after a $q$-swap operation has been done we arrive at a cycle of
$q$ nodes (including the one we are coming from) connected via edges occupied
with probability $\phi_1$ (see Figure~\ref{fig:branching_qswap}). When edges
are converted into singlets, the accessible nodes of this new $q$-cycle form a
string of length $l$ with probability
\begin{equation*}
  \left\{ \begin{array}{ll}
    \phi_1^q                  & \textrm{for $l=q$,} \\
    q\phi_1^{q-1}(1-\phi_1)   & \textrm{for $l=q-1$,} \\
    (l+1)\phi_1^l(1-\phi_1)^2 & \textrm{for $l\le q-2$.}
  \end{array} \right.
\end{equation*}
For $l\le q-2$, $l$ new components emerge, with total size (including all
the vertices in the cycle, except the starting one) probability generated by
$[xg_r(\tilde{h}_R(x))]^l$. For $l=q-1$ and $l=q$, $q-1$ components emerge,
again with total size probability generated by $[xg_r(\tilde{h}_R(x))]^{q-1}$.
The total size of such cycle and its emerging components is then generated by
\begin{eqnarray}
  C_q(x)
  &=& \sum_{l=0}^{q-2} (l+1) \phi_1^l (1-\phi_1)^2 \left[ x g_r(\tilde{h}_R(x)) \right]^l\nonumber
  \\
  &&+ \left[ q\phi_1^{q-1} (1-\phi_1) + \phi_1^q \right] \left[ x g_r(\tilde{h}_R(x))
  \right]^{q-1}.
  \label{eq:Cq}
\end{eqnarray}
Therefore, the new $\tilde{h}_R(x)$ is of the same form of
Eq.~(\ref{eq:component_size_edge_gf}) plus a term $\tilde{h}_{R,q}(x)$ for
each $q$-swap:
\begin{align}
  \tilde{h}_R(x)
  &= 1 - \phi_2 + \phi_2 x g_r ( \tilde{h}_R(x) )
  + \sum_{q\ge2} \Pi_q \tilde{h}_{R,q}(x)
  \label{eq:new_component_size_edge_gf} \\
  \tilde{h}_{R,q}(x)
  &= 
  r_{q-1}
  \left[
  (\phi_2-1)
  - \phi_2 x \left( \tilde{h}_R(x) \right)^{q-1}
  + C_q(x)
  \right].\nonumber
\end{align}
At this stage we can already calculate $\tilde{\phi}_1^*$ as the
smallest value of $\phi_1$ for which there exists a positive solution
$\tilde{u}=\tilde{h}_R(1)<1$ to \eqref{eq:new_component_size_edge_gf} at
$x=1$. It is easy to convince oneself that each separate contribution
$\tilde{h}_{R,q}(1)$ either increases or lowers the percolation threshold and
therefore for the optimal strategy each $\Pi_q$ is either 0 or 1.

For the new $\tilde{h}_P(x)$ we need to consider that not all nodes of degree
$q$ are suitable targets of $q$-swaps, since they cannot be performed on
adjacent nodes. Therefore, given a node of degree $q$ there is a probability
$\eta_q$ that a $q$-swap can be performed on it. If the $q$-swap is performed
on a node, then it changes its degree from $q$ to zero and hence
\begin{equation}
  \tilde{h}_P(x)
  = xg_p[\tilde{h}_R(x)] + x\sum_{q\ge2} \Pi_q\eta_q p_q
  \{1-[\tilde{h}_R(x)]^q\}.
  \label{eq:new_component_size_vertex_gf}
\end{equation}
By using the solution $\tilde{u}=\tilde{h}_R(1)$ here, we can obtain the size
of the giant connected component, $\tilde{S} = 1-\tilde{h}_P(1)$. This gives
the probability $\hat{S}^2$ that two distant nodes are connected by a path of
singlets.

\begin{figure}[t]
  \begin{center}
    \includegraphics[width=.45\textwidth]{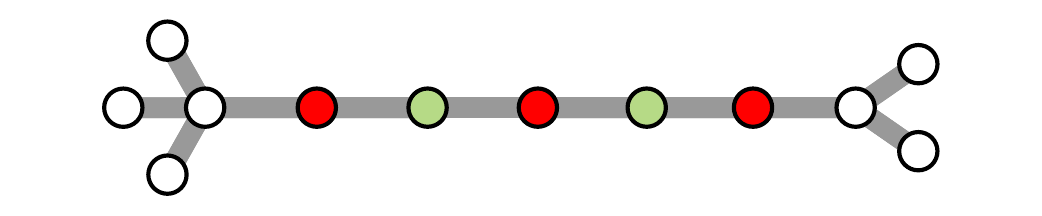}
    \\
    \includegraphics[width=.45\textwidth]{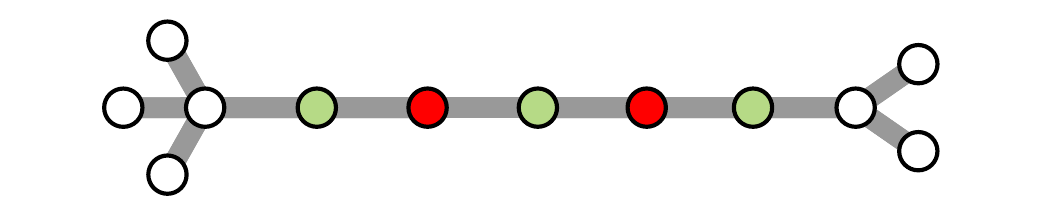}
  \end{center}
  \caption{(Color online) Two clusters of 5 nodes with
  degree 2. Nodes are big circles: empty if their degree is different from
  2, dark grey (red) if they are operated on, light grey (green) if they
  are not. Top: operations are done at nodes 1, 3 and 5, leading to
  $\eta_2^{(\rm max)}$. Bottom: operations are done at nodes 2 and 4, leading
  to $\eta_2^{(\rm min)}$.}
  \label{fig:eta2_net_examples}
\end{figure}
The probability $\eta_q$ depends on which degrees are targets of $q$-swaps
and on how the network is traversed to operate on the nodes. To compute its
value we need to consider maximal clusters consisting of nodes where all
vertices are of any target degree $q$---the border of such clusters is
necessarily made of nodes of degree different from $q$, and hence operations
can be done independently on every cluster. As an example, let us discuss
the simplest case of only performing 2-swaps. Starting from a random vertex
of degree 2, we find a cluster of vertices of same degree 2 whose size is
$s$ with probability $s(1-r_1)^2r_1^{s-1}$. By acting on a node, and then on
every second node, there are two possible values for the number of operations
done in each cluster, $\lceil s/2\rceil$ and $\lfloor s/2\rfloor$ (Figure
\ref{fig:eta2_net_examples}), which coincide for $s$ even. This gives a
maximum and minimum value for $\eta_2$,
\begin{align}
  \eta_2^{(\rm max)}
  &= \sum_{s\ge1} s(1-r_1)^2r_1^{s-1}\frac{\lceil s/2\rceil}{s}
  = \frac{1}{1+r_1}
  \label{eq:eta2_max}
  \\
  \eta_2^{(\rm min)}
  &= (1-r_1)^2+\sum_{s\ge2} s(1-r_1)^2r_1^{s-1}\frac{\lfloor s/2\rfloor}{s}
  \notag \\
  &= \frac{1-(1-r_1)r_1^2}{1+r_1}.
  \label{eq:eta2_min}
\end{align}
Note that for clusters of size $s=1$, an operation is always done. When
operations are performed starting from a random vertex in each cluster of
vertices with degree 2, one needs to take into account the number of vertices
$s$ and $t$ at odd and even (including zero) distance from the first vertex:
operations will be performed on a fraction $t/(t+s)$ of the cluster. The
probability $\xi(s,t)$ of starting in a vertex of degree 2 such that it has
$s$ neighbors of degree 2 at odd distance and $t$ at even distance is
\begin{equation}
  \xi(s,t) = \binom{2}{1+s-t}(1-r_1)^2r_1^{s+t-1}t
  \label{eq:xi2}
\end{equation}
if $|s-t|\le1$ and 0 otherwise. For general $q$, this probability can be
found by generating functions similar to the ones described before, see
Appendix~\ref{sec:appetaq} for more details. Given the probability $\xi(s,t)$,
then the value for $\eta_2$ when operations are started at each cluster of
target vertices is
\begin{equation}
  \eta_2^{(\rm rand)}
  = \sum_{s,t} \frac{t}{t+s}\xi(s,t)
  = \frac{r_1+(1-r_1)^2 \atanh(r_1)}{2r_1}.
  \label{eq:eta2_rand}
\end{equation}

\begin{figure}[t]
  \begin{center}
    \includegraphics[width=.45\textwidth]{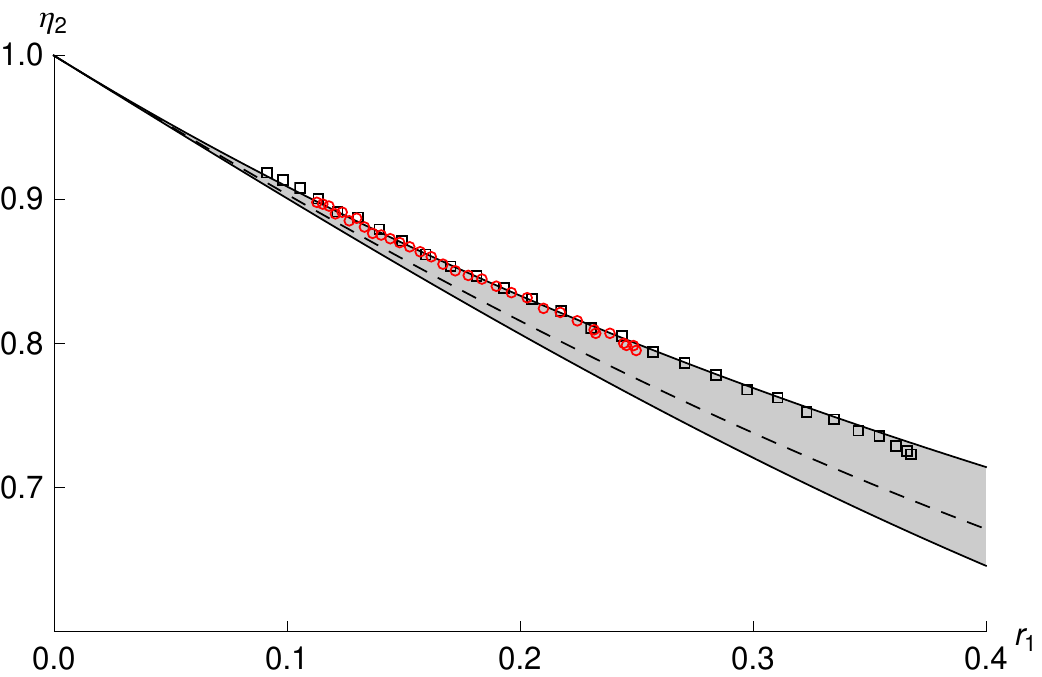}
  \end{center}
  \caption{(Color online) Probability $\eta_2$ of performing a 2-swap,
  given a vertex of degree 2. Upper and lower lines correspond to Eqs.\
  (\ref{eq:eta2_max}) and (\ref{eq:eta2_min}) respectively, squares to Erd\H
  os--R\'enyi network simulations, circles to scale free network simulations.
  All the simulations were performed with $N=10^6$ and Bread First Search
  traversal of the graph.}
  \label{fig:eta2}
\end{figure}

Figure~\ref{fig:eta2} show $\eta_2^{(\rm max)}$, $\eta_2^{(\rm min)}$ and
$\eta_2^{(\rm rand)}$ together with numerical simulations performing 2-swaps
by traversing the graph with a Breadth First Search, as described in the
following section. The numerical values for $\eta_2$ are close to the maximum
value because it is much more likely that the traversal of graph started
outside most of the degree 2 clusters (e.g., arriving through one of the white
nodes in Figure \ref{fig:eta2_net_examples}), thus performing the maximum
number of operations in them.

\subsection{Network examples and simulations}

Here we present some examples of entanglement percolation in the networks
described in Section~\ref{sec:net_examples}, and we provide analytic solutions
for paradigmatic cases. To check these results and extend them to correlated
and real world networks we have performed computer simulations with various
networks models. Graphs with uncorrelated degree distribution are relatively
easy to generate \cite{Newman2001}. First, a set of $N$ numbers $\{k_i\}$
randomly chosen to follow the desired degree distribution is generated, so
each vertex $i$ has $k_i$ stubs or ``half edges'' associated with it. If the
sum $\sum_i k_i$ is odd, a new set is generated until an even sum is obtained,
so all stubs can be joined. Then pairs of stubs are selected randomly and
joined to form edges until there are no stubs left. In our simulations we did
not allow self-loops or multiple edges. The quantum preprocessing is done by
traversing all the graph with a Breadth First Search (BFS), which starts at
a random root vertex and explores all the neighboring nodes at distance 1,
2\dots, in order, until all the vertices in the component have been visited.
After that, another BFS is done starting from a random unexplored vertex in
another component, until all components have been examined. At each discovered
vertex, the local structure is changed from a $q$-star to a $q$-cycle if the
vertex degree is one of the target degrees and if non of the edges in the star
have already been used.

\begin{figure}[t]
  \begin{center}
    \includegraphics[width=.22\textwidth]{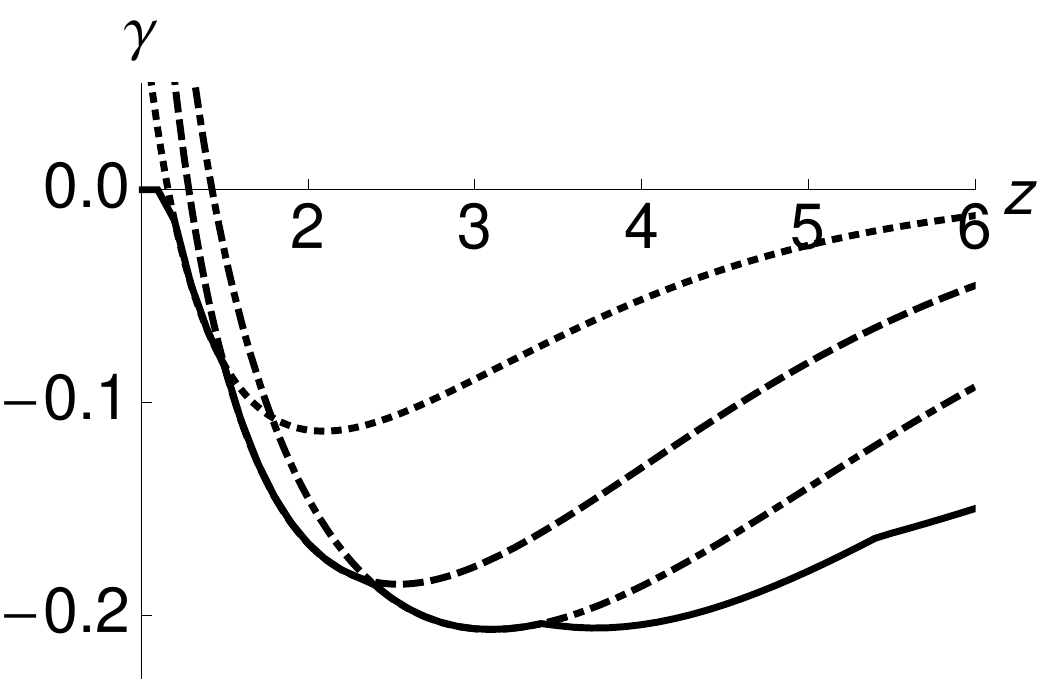}
    \includegraphics[width=.22\textwidth]{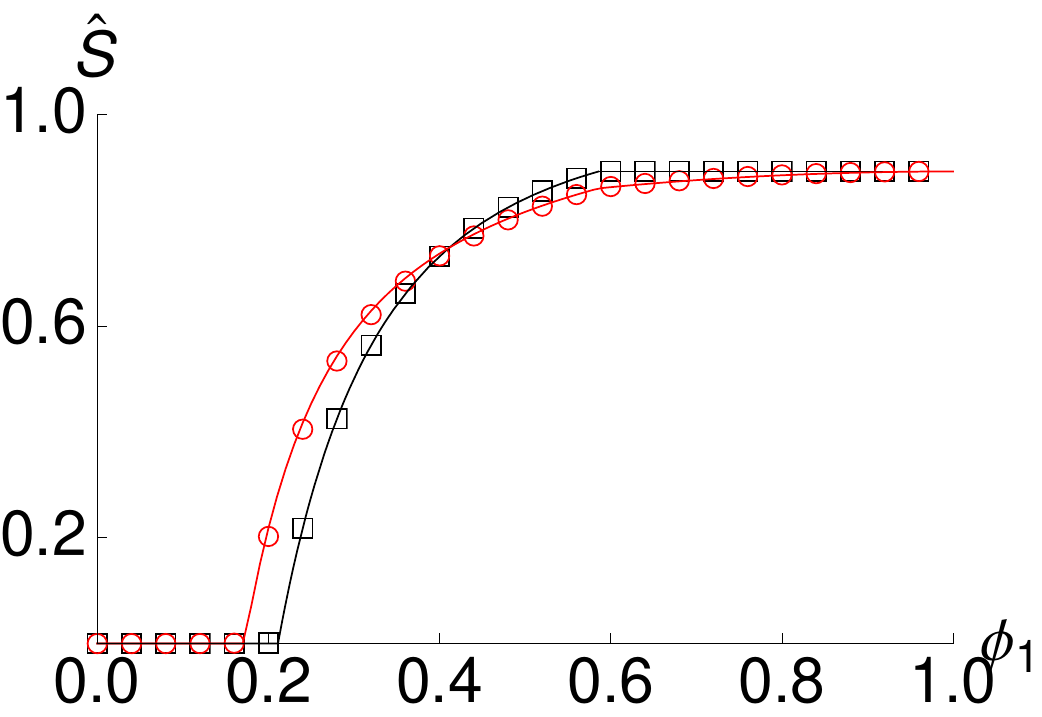}
  \end{center}
  \caption{(Color online) Erd\H os--R\'enyi network. (a) Gain $\gamma$ as a
  function of the mean degree $z$ after 2-swap (dotted line), 2,3-swap (dashed
  line), 2,3,4-swap (dot-dashed line) and optimal $q$-swaps (solid line). (b)
  Normalized size $\hat{S}$ of the GCC as a function of $\phi_1$ for $z=2.5$,
  before (squares) and after (circles) 2,3-swap, $N=10^6$.}
  \label{fig:er}
\end{figure}

\begin{figure}[t]
  \begin{center}
    \includegraphics[width=.22\textwidth]{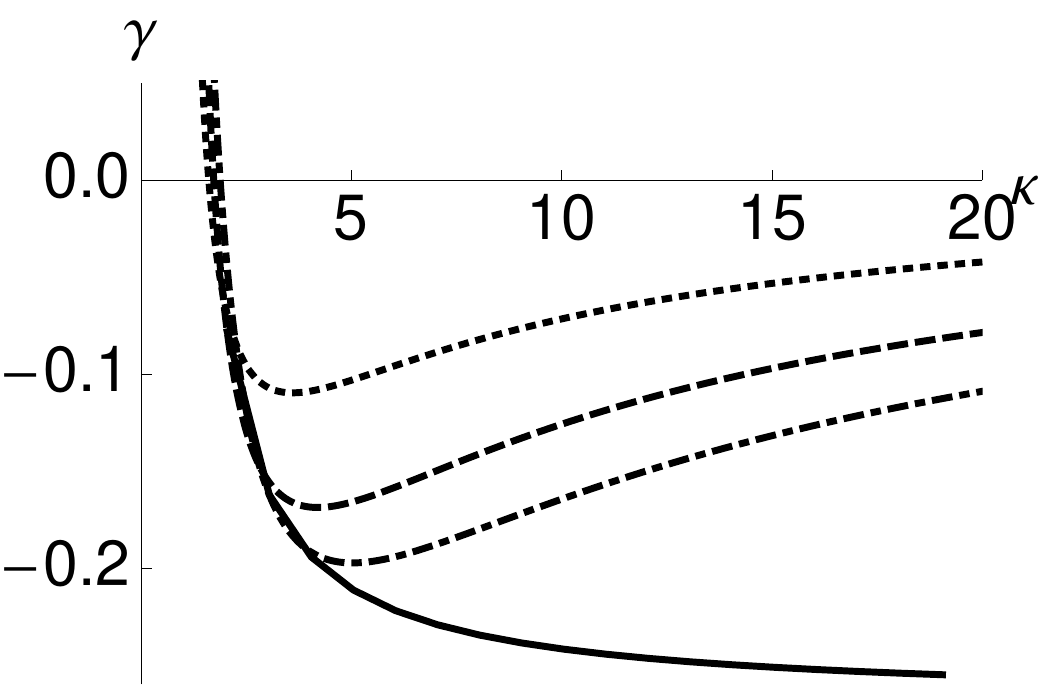}
    \includegraphics[width=.22\textwidth]{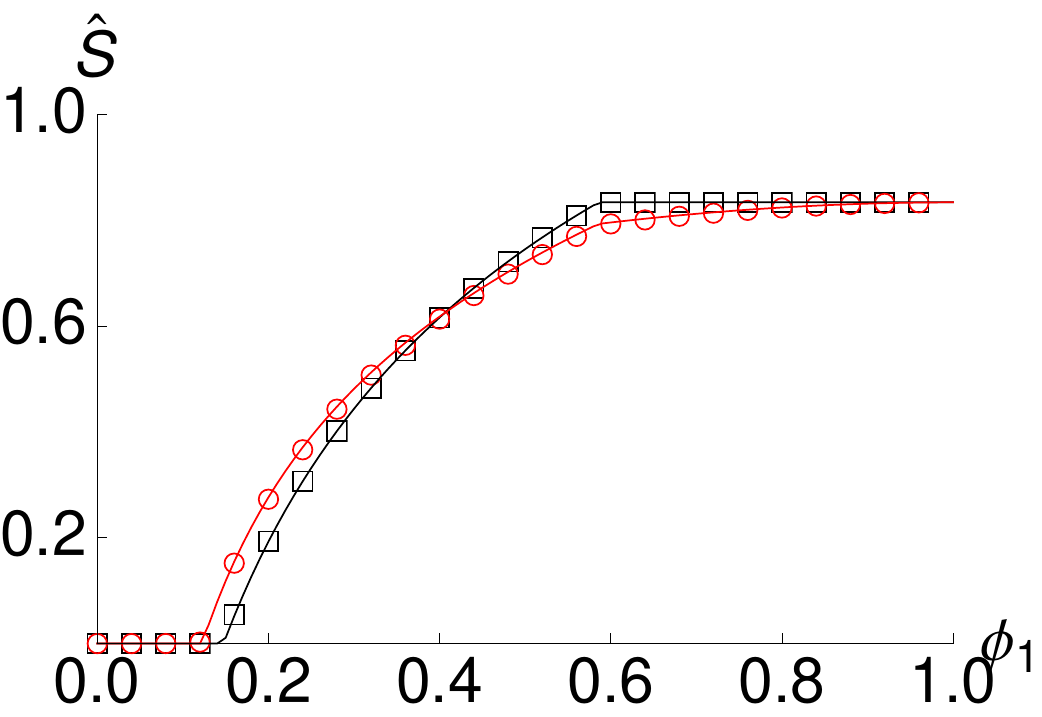}
  \end{center}
  \caption{(Color online) Scale free network, $\tau=1$. (a) Gain $\gamma$ as
  a function of the cutoff $k$ after 2-swap (dotted line), 2,3-swap (dashed
  line), 2,3,4-swap (dot-dashed line) and optimal $q$-swaps (solid line). (b)
  Normalized size $\hat{S}$ of the GCC as a function of $\phi_1$ for $\kappa=4$, before
  (squares) and after (circles) 2,3-swap, $N=10^6$.}
  \label{fig:plec}
\end{figure}

For the Bethe lattice with coordination number $q$, $g_p(x)=x^q$ and
$g_r(x)=x^{q-1}$. In this network the phase transition occurs at
$\phi_2=(q-1)^{-1}$ and
\[
(q-1)^{-1}=(1-\phi_1)^{-1}\{2\phi_1+\phi_1^q[\phi_1(q-1)-(q+1)]\}
\]
after $q$-swap is applied. Therefore, $q$-swap gives always a better threshold
except for the special case $q=2$ of an infinite 1D chain, where the
probability decays exponentially with the distance.

In the Erd\H os--R\'enyi network, before any transformation the threshold is
given by $\phi_2=1/z$. After, e.g., the 2-swap and 3-swap operations, the
thresholds are, respectively,
\begin{align}
  \frac{1}{z} &= \phi_2+e^{-z}[-\phi_2+z(2\phi_1-\phi_1^2)] \\
  \intertext{and}
  \frac{1}{z} &= \phi_2+ze^{-z}[-\phi_2+z(1+\phi_1-\phi_1^2)].
\end{align}
Figure~\ref{fig:er} shows the evolution of the giant connected component
before and after the transformations, with perfect agreement between
analytical and numerical results. Figure~\ref{fig:er} also shows the gain
$\gamma=(\tilde{\phi}_{1}^*-\phi_{1}^*)/\phi_1^*$ in the percolation
threshold, which in some situations is higher than 20\%. The performance of
different $q$-swaps depends on the mean degree $z$, usually improving the
threshold those operations which act on nodes whose degree is around $z$.
Figure~\ref{fig:plec} show similar results for the giant connected component
evolution and the gain in scale free networks with $\tau=1$. In this case the
gain can be of around 25\%.

\begin{figure}[t]
  \begin{center}
    \includegraphics[width=.22\textwidth]{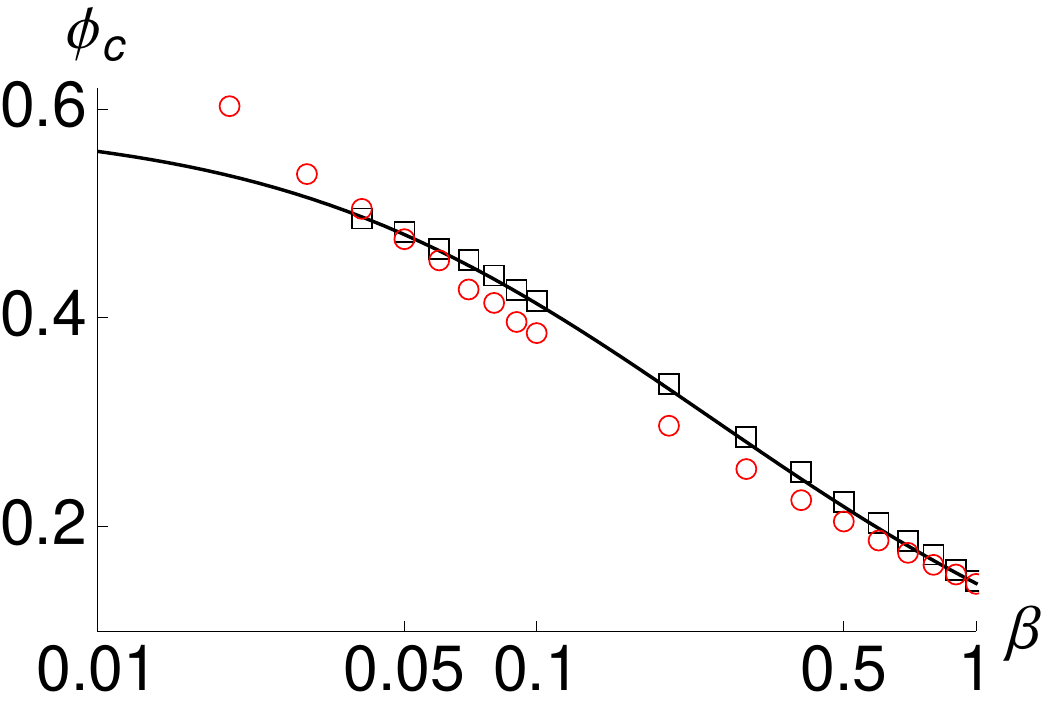}
    \includegraphics[width=.22\textwidth]{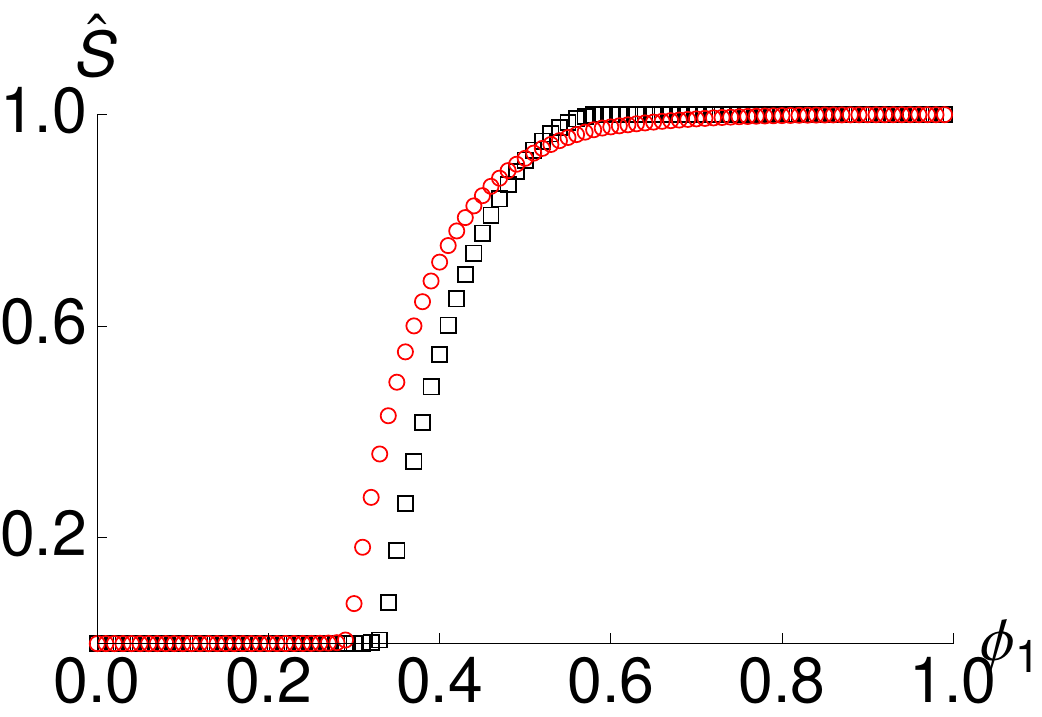}
  \end{center}
  \caption{(Color online) Watts--Strogatz network. (a) Percolation threshold
  $\phi_1^*$ as a function of shortcut probability $\beta$ before (squares)
  and after (circles) 2-swap. The solid line is the analytic result from
  Ref.~\cite{Moore2000a}. (b) Normalized size $\hat{S}$ of the GCC as a
  function of $\phi_1$ for $\beta=0.2$, before (squares) and after (circles)
  2-swap, $N=10^6$.}
  \label{fig:swas}
\end{figure}

\begin{figure}[t]
  \begin{center}
    \includegraphics[width=.42\textwidth]{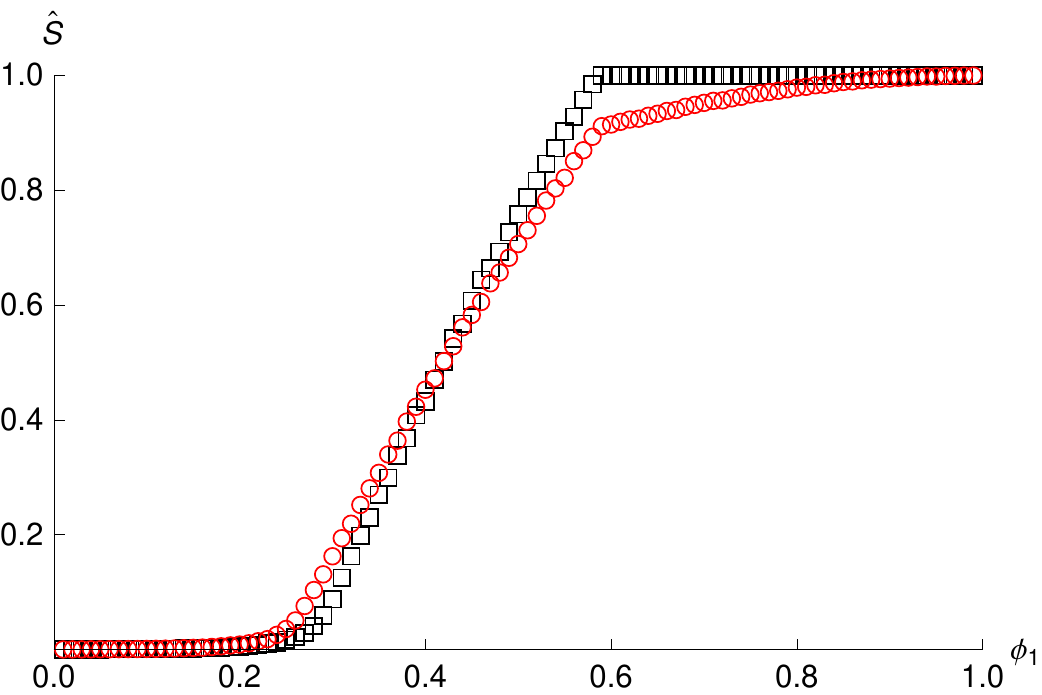}
  \end{center}
  \caption{(Color online) World Wide Web network \cite{Albert1999}, with a
  cutoff at $k=15$, before (squares) and after (circles) 2,3-swap.}
  \label{fig:www}
\end{figure}

For the Watts--Strogatz model, which is correlated, and the World Wide Web
network, the above approach is not valid because the tree-like assumption
does not hold. However, numerical simulations show that $q$-swaps can also
provide an improvement in the percolation threshold,
$\tilde{\phi}_1^*<\phi_1^*$.
Figure~\ref{fig:swas} shows the threshold probability for the Watts--Strogatz
before and after 2-swap and the size of the giant connected component.
Figure~\ref{fig:www} shows the size of the giant connected component for the
World Wide Web.

Note that, in general, it may be counterproductive to perform $q$-swaps. In
the above figures we see that for some values of $\phi_1$ the giant connected
component fraction $S$ without preprocesing is larger than $\hat{S}$. This
often happens around $\phi_1=2-\sqrt{2}$. This is precisely the point where
the edges in the unmodified network can be directly converted into singlets
with $\phi_2=1$, i.e. all connections become ideal channels and $S$ attains
its maximal value $S=S_{1}$. Obviously at this stage any preprocessing cannot
further increase the size of the connected component, and it will most likely
decrease it.

\section{Mixed state networks}

A network with nodes connected by pure states is an abstraction that gives
insight into the possibilities of long-distance entanglement in complex
networks, enabling perfect teleportation between distant parties when at least
a path of maximally entangled states is created. In general, however, states
connecting two neighbors are noisy and need to be described by mixed states.
In this situation, the optimal fidelity of teleportation $f$ is directly
related to the maximal singlet fraction $F$ \cite{Horodecki1999} by
\begin{equation}
  f = \frac{Fd+1}{d+1},
  \label{eq:teleportation_fidelity}
\end{equation}
where $d$ is the dimension of each part of the bipartite state and $F$ is
defined as the maximal overlap of a state $\rho$ with a maximally entangled
state $\ket{\Psi}=\frac{1}{\sqrt{d}}\sum_i\ket{ii}$,
\begin{equation}
  F(\rho) = \max_{\ket{\Psi}}\bra{\Psi}\rho\ket{\Psi}.
  \label{eq:singlet_fraction}
\end{equation}
Entanglement percolation in the mixed state scenario has already been
addressed in regular lattices with connections consisting singlet that have
suffered an amplitude damping \cite{Broadfoot2009, Broadfoot2010}. There,
an hybrid swapping strategy is proposed for this type of connections,
which could also be used to build a mixed state $q$-swap to act on complex
networks with at least four states per edge. In another approach, Perseguers
gives a fidelity threshold for the links above which long-distance quantum
communication in the presence of noise is possible for an infinite cubic
lattice \cite{Perseguers2010}.

\begin{figure}[tb]
  \begin{center}
    \includegraphics[width=.22\textwidth]{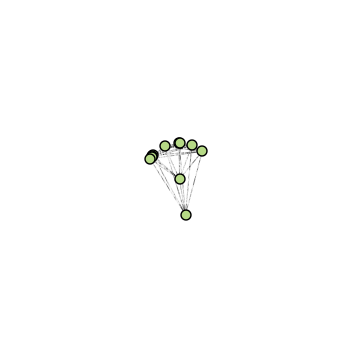}
    \includegraphics[width=.22\textwidth]{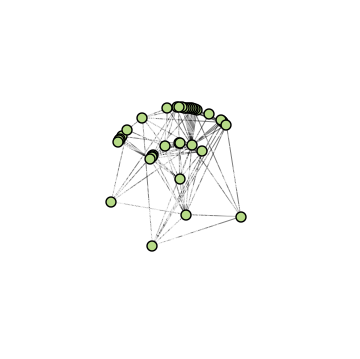}
    \\
    \includegraphics[width=.22\textwidth]{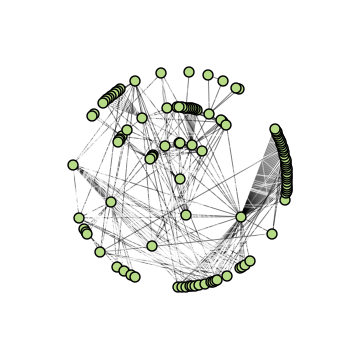}
    \includegraphics[width=.22\textwidth]{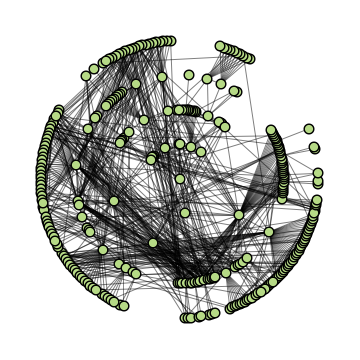}
  \end{center}
  \caption{(Color online) For a limited path length, cluster growth depends on
  the network topology. Here, clusters of limited path length $l=1,2,3,4$ in
  the OpenPGP Web of Trust, with sizes $11, 37, 115, 286$.}
  \label{fig:pgp_wot_evolution}
\end{figure}

Communication in noisy networks can be considered from another perspective.
The noise in the connections fixes a limit $l$ in the maximum number of nodes
through which the information can be repeated before it becomes too corrupted
\cite{Dur1999}. In this limited path length scenario, the total number of
vertices that a given node can communicate to also depends strongly on the
structure of the communication network (Figure \ref{fig:pgp_wot_evolution}).
This is related to the average path length $l_{\rm av}$: the length of the
shortest path averaged over all possible pairs of nodes. All nodes within this
distance constitute a significant fraction of the network. Therefore, for a
path length limit $l$ above the average $l_{\rm av}$, communication will be
possible among an important number of nodes. Since the limiting $l$ is finite,
the giant connected component appears only in models where $l_{\rm av}$ is
also finite. In general, this only happens if the network size is finite too.
The question then is whether a small $l$ will suffice to cover a significant
fraction of the network. In finite $d$-dimensional networks, the average path
length scales as $l_{\rm av} \sim N^{1/d}$. However, the average path length
of many complex networks scales \emph{logarithmically} with the size of the
network. This property is known as the small world effect, and appears also
in many real world communication networks such as Internet. In this case, to
access a significant fraction of nodes, only a small number of edges need to
be traversed. Small world models are therefore the first candidates, where
loses by noise can be balanced by a short path length.

The problem of \emph{limited path percolation} was also addressed in a
different approach by L\'opez \etal\ \cite{Lopez2007}. In their model,
they calculate the percolation phase transition under the assumption that
communication is only effective if the new minimum path length between two
nodes do not exceed a multiple of the original path length between them. Thus,
in their study the limitation in the path length comes from the topology of
the network and not from the nature of channels connecting nodes, which fixes
a constant limit of nodes through which the information can be repeated.

Here, we are interested in the number of nodes that can exchange quantum
information with a given node for some fixed minimum fidelity, or similarly
with what probability two random nodes can reliably communicate between them.
We will consider a similar scenario as in the previous sections, but replacing
pure-state connections with generic entangled mixed states. Here, no quantum
preprocessing will be possible. However, we will find that the complex network
structure (in particular the small world effect) allows to interconnect a
large number of nodes using the standard entanglement percolation strategy.
We start by doing some numerical simulations and then derive the generating
functions for limited path percolation and compute the limited average size in
non-correlated networks and the Watts--Strogatz model.

\subsection{Network simulations}

We begin by simulating different models of networks. For simplicity we
consider that edges hold a single copy of a two qubits state $\rho_F$ with
maximum singlet fidelity $F>1/2$ so that the classical limit of $f=2/3$ in
the teleportation fidelity can be exceeded. Note that, as long as $\rho_F$ is
entangled, this limit can be achieved even if $F<1/2$ by locally increasing
the singlet fidelity through trace-preserving local operations and classical
communication (LOCC) \cite{Verstraete2003}. By applying random bilateral
rotations, $\rho_F$ can be brought into a Werner state,
\begin{multline}
  \rho_F
  = F \ketbra{\Psi^-}{\Psi^-} + \frac{1-F}{3} \ketbra{\Psi^+}{\Psi^+} \\
  + \frac{1-F}{3} \ketbra{\Phi^-}{\Phi^-} + \frac{1-F}{3}
  \ketbra{\Phi^+}{\Phi^+},
  \label{eq:werner_state}
\end{multline}
which has the same singlet fidelity $F$. This state can also be written as
\begin{equation}
  \rho_\alpha = \alpha \ketbra{\Psi^-}{\Psi^-} + (1-\alpha) \frac{\openone}{4},
  \label{eq:depolarized_state}
\end{equation}
with $\alpha=(4F-1)/3$. It can be interpreted as the result of transmitting a
pure singlet through a depolarizing channel. Hence, when a state is teleported
through $l$ of such edges \cite{Briegel1998, Dur1999, SenDe2005}, its fidelity
is 1 with probability $\alpha^l$ and $1/2$ otherwise, so its final fidelity is
$f_l=(1+\alpha^l)/2$. This fidelity decreases exponentially with the distance
$l$ and makes such communication scheme useless in networks such as linear
chains or regular lattices, where the typical distance between two nodes
scales as the size of the network. However, as we discussed, the typical
distance in many complex networks scales only logarithmically. The maximum
distance $l$ that information can travel is fixed by the minimum fidelity
$f_{\rm min}$ required at the end point and by the purity $\alpha$ of the
channels,
\begin{equation}
  l = \left\lfloor \frac{\ln (2f_{\rm min}-1)}{\ln\alpha} \right\rfloor.
  \label{eq:lmax}
\end{equation}
This means that, even if there exists a path between a sender and a receiver
in a network, it will only be useful if the length of this path is below a
certain threshold.

\begin{figure}[t]
  \begin{center}
    \includegraphics[width=.45\textwidth]{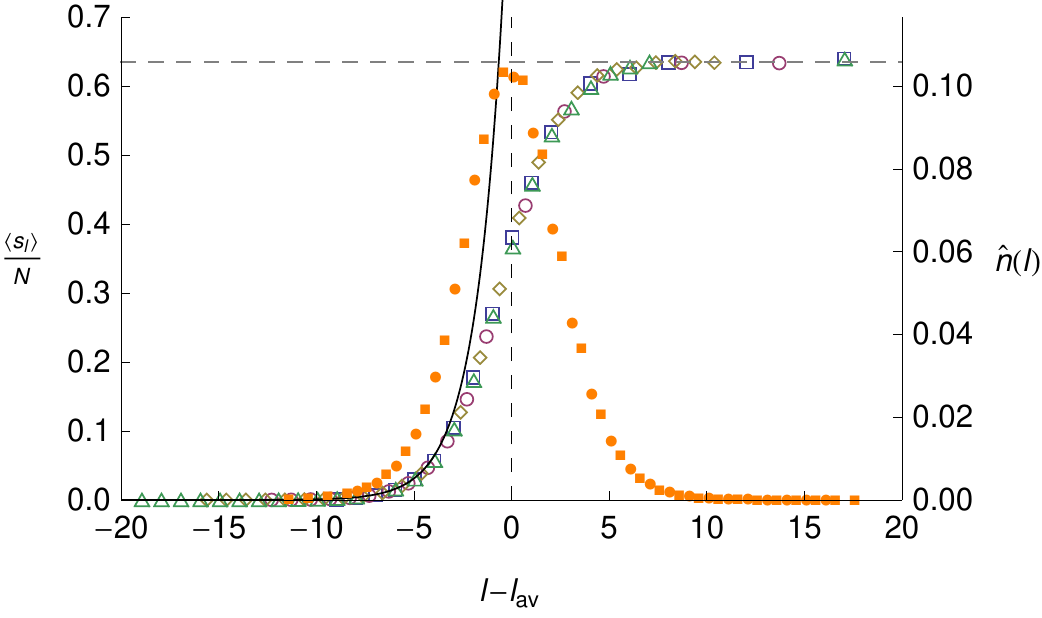}
  \end{center}
  \caption{(Color online) Normalized $l$-limited average component size
  $\langle s_l\rangle/N$ as a function of $l-l_{\rm av}$ for the Erd\H
  os--R\'enyi network with $k=2$ and network sizes $N=10^3, 10^4, 10^5, 10^6$
  (squares, circles, diamonds and triangles). Superposed filled markers
  (in orange) is the shape of the path length distribution normalized with the
  total number of possible vertex pairs, $\hat{n}(l)=2n(l)/N(N-1)$. Solid
  black line is Eq.~(\ref{eq:average_limited_size}), horizontal dashed line is
  the square of the GCC at $\phi_1=1$, see Eq.~(\ref{eq:erS}).}
  \label{fig:er_lcp_scaling}
\end{figure}

\begin{figure}[t]
  \begin{center}
    \includegraphics[width=.45\textwidth]{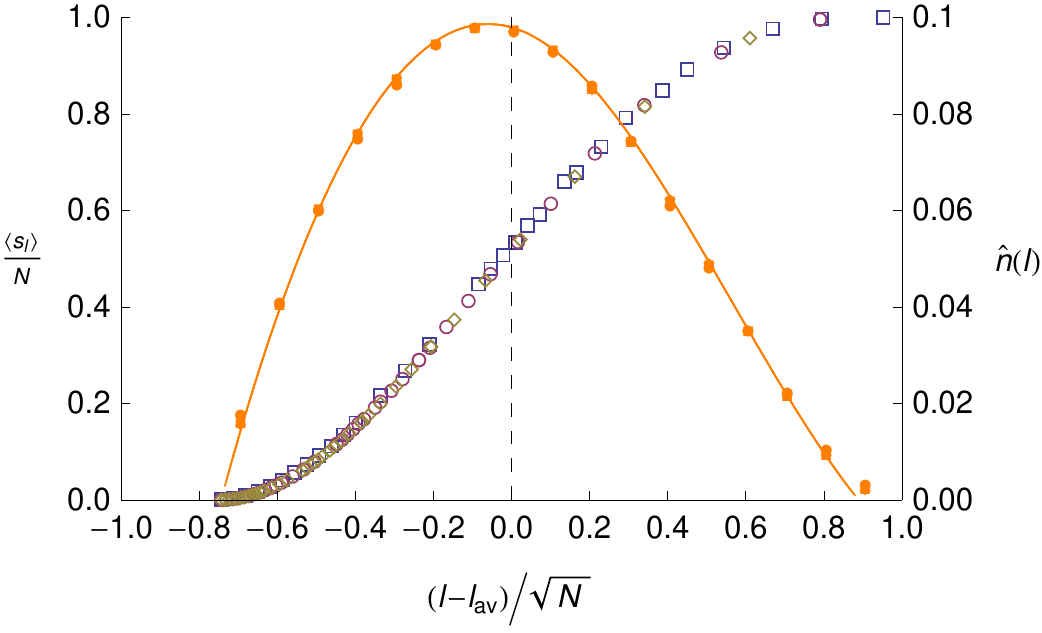}
  \end{center}
  \caption{(Color online) Normalized $l$-limited average component size
  $\langle s_l\rangle/N$ as a function of $(l-l_{\rm av})N^{-1/2}$ for the
  Honeycomb 2-dimensional network with network sizes $N=1014,5046,10086$
  (squares, circles and diamonds). Superposed filled markers (in orange) is
  the normalized histogram of the path length distribution.}
  \label{fig:hc_lcp_scaling}
\end{figure}

We performed extensive simulations of networks where neighboring nodes
share a state (\ref{eq:depolarized_state}), and considered the classical
limit as the minimum required fidelity, $f_{\min}=2/3$. For small networks
($N\lesssim10^4$) we performed the calculations over several network
realizations and then averaged the results. For bigger networks, a single
network realization is usually enough due to the self-averaging. The
$l$-limited average cluster size $\langle s_l\rangle$ is a specially relevant
parameter, which amounts to the probability that two nodes can communicate
with fidelity $f>f_{\rm min}$.

We have thus calculated $\langle s_l\rangle$ for different network models
and sizes. In Figure~\ref{fig:er_lcp_scaling} we plot the normalized size
$\langle s_l\rangle/N$ as a function of $l-l_{\rm av}$ for the Erd\H
os--R\'enyi model, with average path length $l_{\rm av}\sim\ln N/\ln z$
\cite{Dorogovtsev2003}. For different network sizes the curves collapse,
supporting a linear $N$-dependence $\langle s_l\rangle\sim N$ for fixed $l$. Similar results
have been recently found for the average number of nodes at exact distance
$l$ from a random central node \cite{Shao2008a}. Regarding the dependence in $l$, our results show that the
average size grows exponentially with $l$ for $l\ll l_{\rm av}$, but deviate
from this behavior when $l$ is close to the average path length, saturating
to the maximum component size shortly after $l_{\rm av}$. This deviation
is due to the depletion of nodes at distance $l>l_{\rm av}$. In the same
Figure~\ref{fig:er_lcp_scaling} we plot the path length distribution, i.e.,
the number of pairs $n(l)$ separated by a distance $l$, normalized by the
total number of pairs $N(N-1)/2$. Again, both curves $N=10^3$ and $N=10^4$
collapse, thus supporting a dependence $n(l)\sim N^2$. We also found similar
results for the scale free and the Watts--Strogatz models. It is interesting to
note that, while $l_{\rm av}$ grows with the size of the network, the width
of the path length distribution remains constant. Thus, for large networks
a small increase in $\alpha$ near $l_{\rm av}$ leads to an abrupt change in
$\langle s_l/N\rangle$. This is in stark contrast to regular lattices, where
both the mean and the width scale as $N^{1/d}$. For instance, in Figure \ref{fig:hc_lcp_scaling}
we plot $\langle s_l\rangle/N$ and the path length distribution of the
Honeycomb 2-dimensional lattice as a function of $(l-l_{\rm av})N^{-1/2}$. The
collapse of the curves confirms the $N^{1/d}$ length-scale dependence.

\begin{figure}[t]
  \begin{center}
    \includegraphics[width=.45\textwidth]{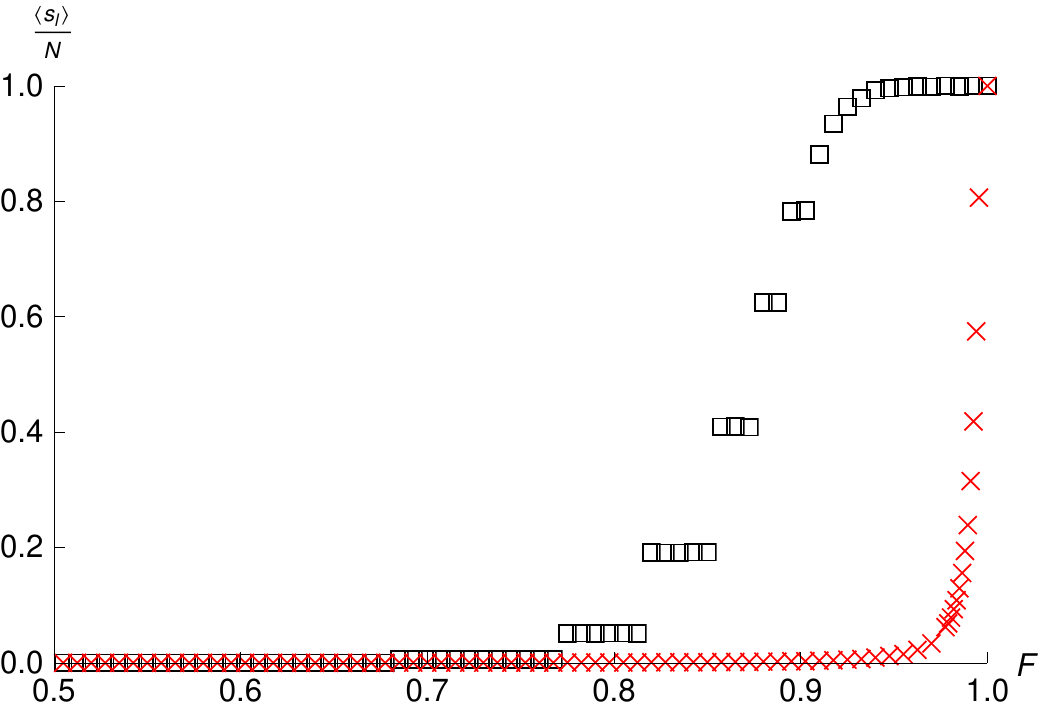}
  \end{center}
  \caption{Normalized $l$-limited average component size $\langle
  s_l\rangle/N$ as a function of singlet fidelity $F$ in the biggest component
  OpenPGP Web of Trust (squares) and a honeycomb 2-dimensional lattice
  (crosses), $N\sim3\cdot10^4$.}
  \label{fig:pgphc_lcp_F}
\end{figure}

As an example of a real work network, we considered the OpenPGP Web of Trust.
Figure~\ref{fig:pgphc_lcp_F} shows the probability that two arbitrary nodes
can communicate with fidelity $f>2/3$ as a function of the
singlet fraction $F$. Again, the comparison with a Honeycomb lattice
of the same size shows that the small world property of the complex networks allows for faithful communication between most of the nodes in the network for reasonable values of the noise, while in regular lattices this is only possible for nearly pure states.

\subsection{Average component size in limited path percolation}

We now proceed to derive the generating functions for the limited path
percolation problem. In this case, we are interested in the distribution of
sizes $s$ of the components that can be reached by only $l$ steps through
edges that are always occupied. As in the non limited case, there are two
different distributions $P_s^{(l)}$ and $R_s^{(l)}$ for the cases where a
random vertex or a random edge are selected. The two corresponding generating
functions, $h_P^{(l)}$ and $h_R^{(l)}$, read as
\begin{align}
  h_P^{(l)}(x)
  &= \left\{ \begin{array}{ll}
    x &\quad \textrm{for } l=0, \\
    x g_p\left[ h_R^{(l-1)}(x) \right] &\quad \textrm{for } l\ge1,
  \end{array} \right.
  \label{eq:lcomponent_size_vertex_gf}
  \\
\intertext{and}
  h_R^{(l)}(x)
  &= \left\{ \begin{array}{ll}
    x &\quad \textrm{for } l=0,\\
    x g_r\left[ h_R^{(l-1)}(x) \right] &\quad \textrm{for } l\ge1.
  \end{array} \right.
  \label{eq:lcomponent_size_edge_gf}
\end{align}
Note that all edges are occupied with probability one. The generalization to a
different occupancy probability is straightforward, but not needed here.

\begin{figure}[t]
  \begin{center}
    \includegraphics[width=.45\textwidth]{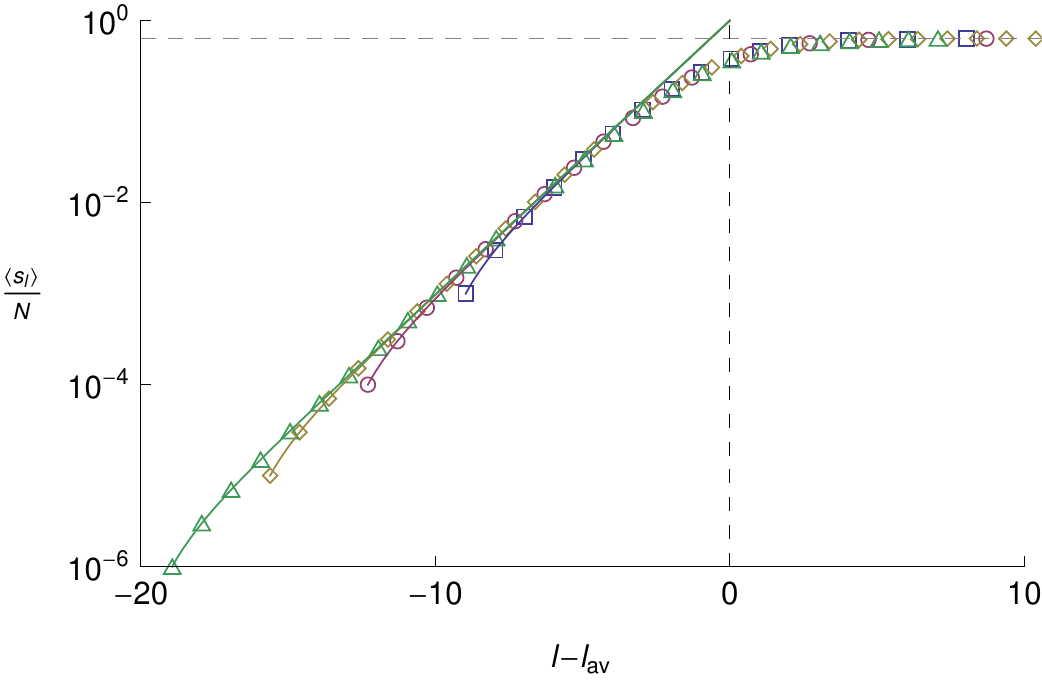}
    \includegraphics[width=.45\textwidth]{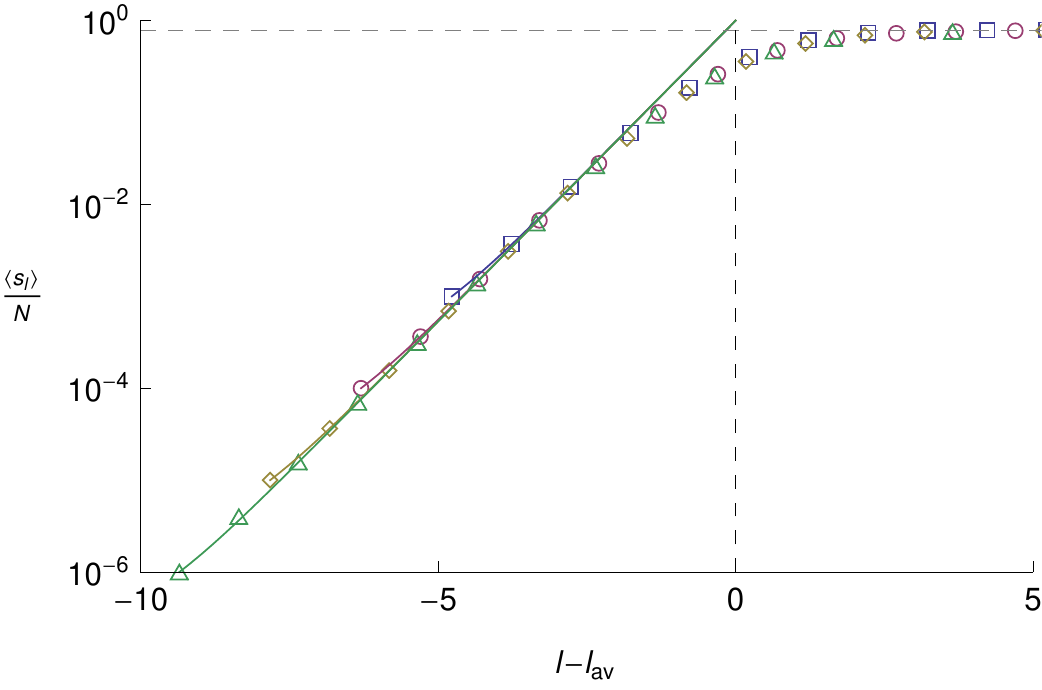}
  \end{center}
  \caption{(Color online) Normalized $l$-limited average component
  size $\langle s_l\rangle/N$ as a function of $l-l_{\rm av}$ for the
  Erd\H os--R\'enyi network (a) and scale free network (b). Points
  are simulation results for network sizes $N=10^3,10^4,10^5,10^6$
  (squares, circles, diamonds, triangles), solid lines correspond to
  Eq.~(\ref{eq:average_limited_size}), horizontal dashed lines are the values
  of $S_1^2$.}
  \label{fig:lcp_scaling_log}
\end{figure}

As before, we are now ready to obtain the $l$-limited average size,
\begin{equation}
  \langle s_l \rangle
  = \left. \frac{\dif h_P^{(l)}(x)}{\dif x} \right|_{x=1} 
  = 1 + g_p\pr(1) h_R^{\prime(l-1)}(1).
\end{equation}
By solving the recurrence equation given by $h_R^{\prime (l)}(1)$ with the
boundary condition $h_R^{\prime (0)}(1) = 1$ we find
\begin{equation}
  \langle s_l \rangle
  = \left\{ \begin{array}{ll}
    1 &\quad \textrm{for } l=0,\\
    1 + g_p\pr(1) \frac{1-(g_r\pr(1))^l}{1-g_r\pr(1)}
    &\quad \textrm{for } l\ge1.
  \end{array} \right.
  \label{eq:average_limited_size}
\end{equation}
This equals to the probability that any two nodes will be able to communicate
with fidelity above $f_{\rm min}$. Figure~\ref{fig:lcp_scaling_log} shows this
result for the Erd\H os--R\'enyi and the scale free model, with very good
agreement between theoretical and numerical results below $l_{\rm av}$.

As we discussed above, this exponential growth of $\langle s_l\rangle$
is valid for $l$ well below $l_{\rm av}$. Our numerical simulations show
that the validity of this approximation can be extended to values near
$l_{\rm av}$. Figure \ref{fig:er_lcp_scaling} shows that the path length
distribution is very peaked around $l_{\rm av}$, and its width is independent
of $N$. This implies, on one hand that our analytical approach holds true
for values of $l$ that fall out of this finite width (approaching from
below)---see Fig. \ref{fig:lcp_scaling_log}. On the other hand, the finite
width implies that if $l$ is a few steps beyond $l_{\rm av}$ then most of
the nodes in the components will be reached before the limit distance is
attained. In this situation, Equations (\ref{eq:lcomponent_size_vertex_gf})
and (\ref{eq:lcomponent_size_edge_gf}) approach the non-limited case of
(\ref{eq:component_size_vertex_gf}) and (\ref{eq:component_size_edge_gf})
with $\phi_1=1$, and the size of the giant component $S_l$ tends to the
non-limited size $S_1$. Therefore, for networks with the small world property,
i.e. $l_{\rm av}\sim \log N$, one can interconnect with a threshold fidelity
(say, the classical benchmark $f=2/3$) any arbitrary pair of nodes in the
network provided that the singlet fraction of the edges scales as $ F
=1-\mathcal{O}(1/\log{N}) $ with the size of the network, which is clearly
less stringent than the analogous constrain for $d$-dimensional networks
$F=1-\mathcal{O}(N^{-1/d})$.

We also consider the Watts--Strogatz model presented in Section
\ref{sec:net_examples}, which has a base circular lattice of size $N$
with $\beta N$ randomly added shortcuts. In this case the derivation of
the probability that a random vertex belongs to an $l$-limited cluster of
size $s$, $P_s^{(l)}$, and its generating function $h_P^{(l)}(x)$ uses
the formalism of ``local clusters'' introduced in \cite{Moore2000a}. This
``local clusters'' are clusters in the base lattice (without considering the
shortcuts). For a given $l$, the ``local cluster'' is always of size $2l+1$.
Then, a shortcut at distance $\lambda-1$ from the starting vertex leads to a
(global) cluster of size $s\pr$ with probability $P_{s\pr}^{(l-\lambda)}$. A
random shortcut emerges from the starting vertex with probability $1/N$, from
a vertex at distance $\lambda$ with probability $2/N$, and lies outside the
local cluster with probability $(N-2l+1)/N$. Hence, that shortcut will lead to
a cluster of size $s$ with a probability given by the generating function:
 \begin{equation}
 f(x)=1-\frac{1}{N}\left(2l -1 -  h_P^{(l-1)}(x) -2 \sum_{\lambda=2}^l h_P^{(l-\lambda)}(x)\right)
 \nonumber
 \end{equation}
There are $2\beta N$ shortcut end-points that can similarly contribute to the
total size of the cluster. Recalling that the generating function of the sum
of sizes is the product of the generating function of each size, we find
  \begin{equation}
 h_P^{(l)}(x)= x^{l+1 }f(x)^{2\beta N}
 \end{equation}
where $x^{(l+1)}$ is the generating function corresponding to the starting
``local'' cluster. In the limit of large $N$ this can be simplified to
\begin{equation}
  h_P^{(l)}(x)
  = x^{1+2l} e^{-2\beta \left[ 2l -1- h_P^{(l-1)}(x) - 2\sum_{\lambda=2}^l
  h_P^{(l-\lambda)}(x) \right] }.
  \label{eq:swas_limited_gf}
\end{equation}

\begin{figure}[t]
  \begin{center}
    \includegraphics[width=.45\textwidth]{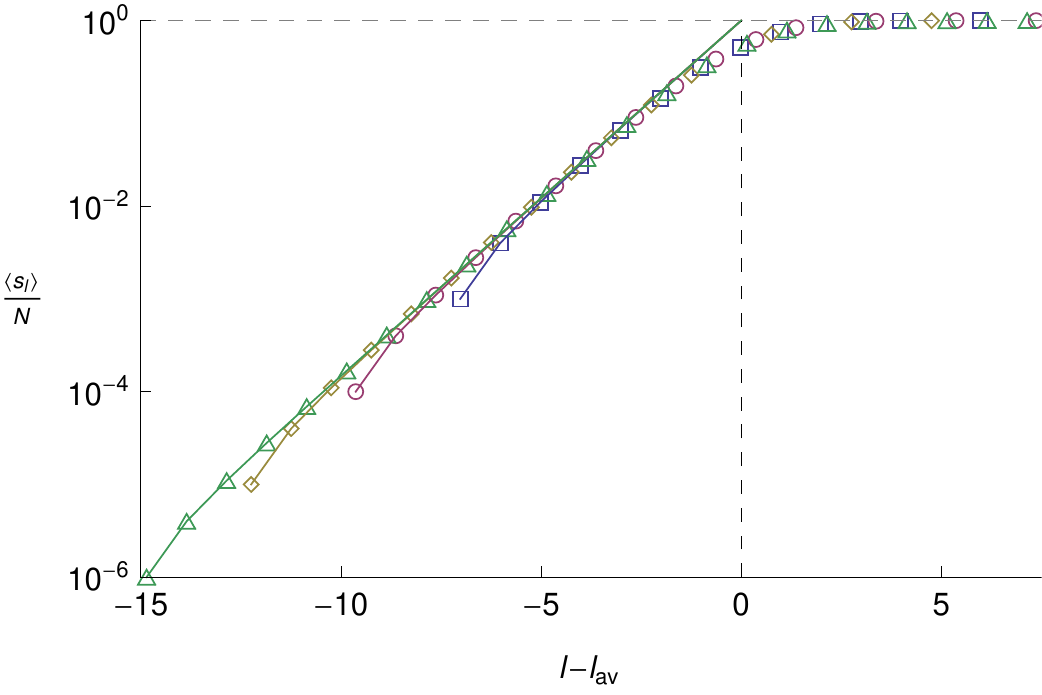}
  \end{center}
  \caption{(Color online) Normalized $l$-limited average component
  size $\langle s_l\rangle/N$ as a function of $l-l_{\rm av}$ for the
  Watts--Strogatz network. Points are simulation results for network sizes
  $N=10^3,10^4,10^5,10^6$ (squares, circles, diamonds, triangles), lines
  correspond to the solution of Eq.~(\ref{eq:swas_limited_average_size}).}
  \label{fig:swas_lcp_scaling_log}
\end{figure}

Again, we can obtain the limited average size by taking the first derivative
at $x=1$. For $l=0$, $\langle s_0\rangle=1$. For $l\ge1$, this results in the
recurrence equation
\begin{align}
  \langle s_l \rangle
  &= 1+2l + 2\beta \left( \langle s_{l-1} \rangle + 2\sum_{\lambda=0}^{l-2}
  \langle s_\lambda \rangle \right) \notag\\
  &= \langle s_{l-1} \rangle + 2 + 2\beta \left( \langle s_{l-1} \rangle +
  \langle s_{l-2} \rangle \right),
  \label{eq:swas_limited_average_size}
\end{align}
which can be exactly solved. Figure~\ref{fig:swas_lcp_scaling_log} shows this
result. We want to stress the fact that from these generating functions,
(\ref{eq:lcomponent_size_vertex_gf}) and (\ref{eq:swas_limited_gf}), one can
also calculate the probability $P_s^{(l)}$ up to any $s$ by solving $s+1$
iterations of them and using Eq.\ (\ref{eq:gf_recover_prob}).

\section{Conclusions}

We have demonstrated that quantum complex networks offer a powerful framework
for entanglement distribution in large systems. Regardless of their intricate
structure, complex networks can be studied by their statistical properties,
which allows to analytically compute some interesting properties and to deal
with them without knowing their exact structure. Here we have considered
entanglement percolation in networks where connections are built on pure,
non-maximal bipartite entangled states, and have studied a local quantum
preprocessing of the network that can significantly decrease the percolation
threshold and therefore allow quantum communication for a lower level of
entanglement. The quantum preprocessing we have proposed is local in two
senses. First, quantum operations are done always on qubits that belong to
the same node. Second, the decision whether or not to perform such operation
depends on the local structure of the network (the degree of the target node
and the status of its neighbors) and on information about general statistical
properties of the network. We have calculated the percolation threshold,
which marks the minimum level of entanglement needed to entangle two distant
nodes with finite probability. We have also computed this probability, which
amounts to the square of the giant connected component in the network. These
results are analytical for networks with uncorrelated degree distribution, and
can be compared to previous results in classical networks, which shows that
the preprocessing can substantially improve communication over such networks
by manipulating its local structure. We have also studied numerically the
Watts--Strogatz small world model and a real world network, and have found a
similar behavior.

In this approach, the links between nodes are pure quantum states. A more
realistic scenario, however, needs to consider noise in the connections. Here
we thus have considered the situation in which such connections are made of
noisy mixed states.

We have shown that in complex networks a direct implementation of the
entanglement percolation strategy, without quantum preprocessing, allows for
faithful quantum communication (above a fixed fidelity threshold) between
a large number of nodes. The noise severely limits the number of steps or
connections through which information is transmitted. However, in complex
networks, one can reach a sizable amount of nodes with a moderately low
number of steps. If the fidelity threshold allows for a path length slightly
higher than the average path length, all nodes in the giant component become
faithfully connected. The path length distribution is peaked at low values
(scaling as $\log N$ in complex networks versus $N^{1/d}$ in $d$-dimensional
lattices), and has finite width (constant in $N$ versus $N^{1/d}$). This
implies that in complex networks a finite fraction of faithfully connected
nodes appears for much smaller limiting path lengths and reaches the giant
component size abruptly. Hence, here the advantage of complex networks
is twofold: the average path length which marks the transition scales
logarithmically with the network size, and the additional steps needed to
reach the non-limited scenario is finite.

We have shown that new phenomena appear if networks and the operations one can
performed on them are governed by the laws of quantum mechanics. This has been
known for regular lattices, but the rich properties of complex networks still
remain widely unexplored in the quantum setting. Our results in percolation,
together with new behavior found in the emergence of subgraphs in quantum
random networks \cite{Perseguers2010b}, are examples of these phenomena.

Our results also contribute to the field of classical complex network. We have
given analytical results for the gain in the percolation thresholds and the
size of the giant component for uncorrelated complex networks that undergo a
set of local inversions (transformation that produce the complement of the
induced subgraph of the target node). The problem at hand of studying how
critical properties of a network can be drastically modified by a given set of
network transformations might be of general interest to other disciplines in
the field. Finally, we have addressed the problem of limited-path percolation
in uncorrelated and small world complex networks.

\begin{acknowledgments}
  We acknowledge financial support from the Spanish ME through FPU grant
  AP2008-03048 (M.~C.); from MICINN through the Ram\'on y Cajal program
  (J.~C.) and projects FIS2008-01236 and QOIT (CONSOLIDER2006-00019); and the
  Generalitat de Catalunya CIRIT, contract 2009SGR985.
\end{acknowledgments}

\appendix

\section{Calculation of \label{sec:appetaq}$\eta_q$}

\begin{figure}[t]
  \begin{center}
    \includegraphics[width=.22\textwidth]{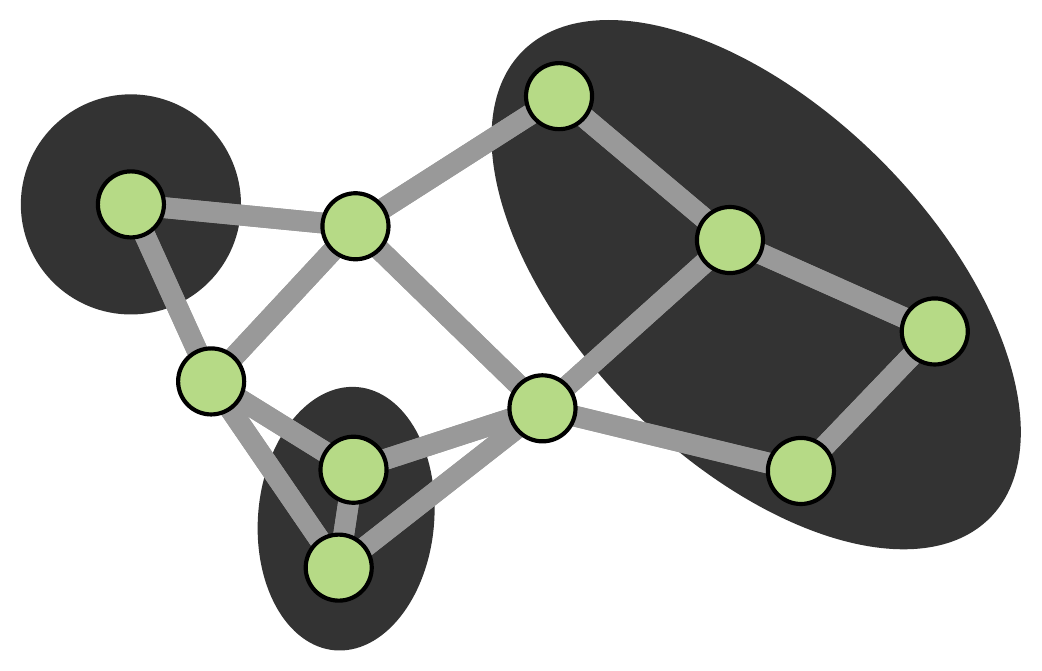}
    \includegraphics[width=.22\textwidth]{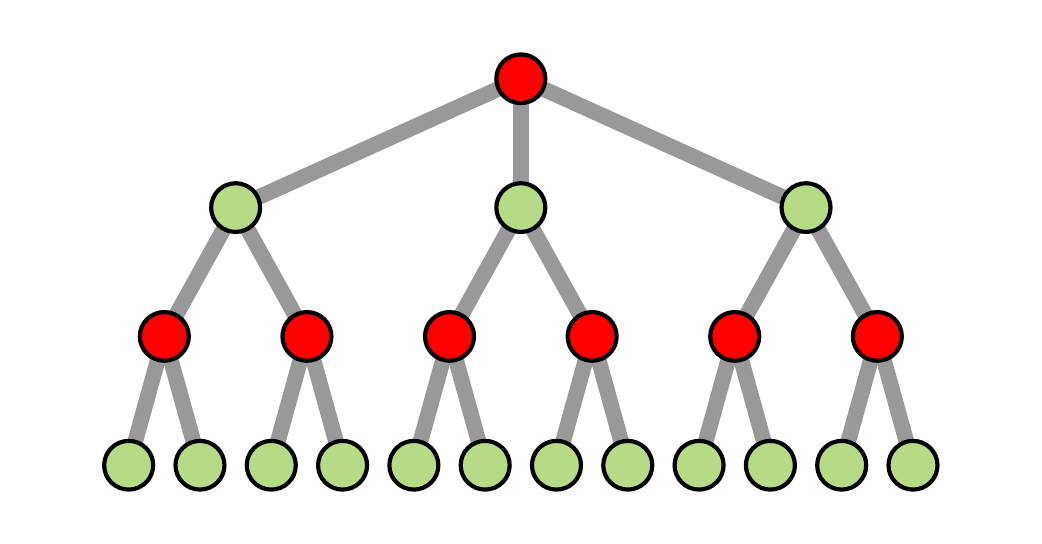}
  \end{center}
  \caption{(Color online) Left: Example of a connected component with three
  clusters (in dark grey) of degree 2 and 3. Right: Branching process in
  $\eta_3^{(\rm rand)}$. 3-swaps are made on dark grey (red) nodes, which are
  the $t$ nodes at even distance from the top one.}
  \label{fig:clusters_eta23_net}
\end{figure}

As we said, the probability $\eta_q$ depends on the target degrees $\{q_i\}$
and on how the network is traversed. By $\eta_q^{(\rm rand)}$ we denote
the probability $\eta_q$ when a $q$-swap is first done in a random vertex
with target degree, and then the cluster of vertices with degree belonging
to $\{q_i\}$ is traversed by a Breadth First Search, performing $q$-swaps
whenever possible (i.e., at every second step). After that, another vertex
with target degree which has not yet been explored is selected, and its
cluster traversed, until all target vertices have been checked. Such
clusters consist of vertices of degree $k\in\{q_i\}$ that are connected
by at least one path whose vertices have also a degree in $\{q_i\}$ and to which no more vertices of degree $k$ can be added.
Figure~\ref{fig:clusters_eta23_net} shows an example of three of such clusters
when the target degrees are 2 and 3. A random vertex of degree $k\in\{q_i\}$
belongs to a cluster with $t$ vertices at even distance (including itself)
and $s$ at odd distance with probability $\xi(s,t)$. In this cluster of size
$s+t$, $t$ $q$-swaps are made. The probability $\eta_q^{(\rm rand)}$ is then
\[
\eta_q^{(\rm rand)} = \sum_{t,s} \frac{t}{s+t} \xi(s,t).
\]

The function generating $\xi(s,t)$ can be computed
similar to Eqs.~(\ref{eq:component_size_edge_gf}) and
(\ref{eq:component_size_vertex_gf}). In this case, it is a function of
two variables: $h_\xi(x,y) = \sum_{s,t\ge0} \xi(s,t) y^sx^t$. Two more
distributions are needed: $S(s,t)$ and $T(s,t)$ are the probabilities of
arriving at a vertex of the given degree (or degrees) which is at an odd or
even distance from the starting vertex, respectively, and which belongs to a
cluster of $s$ extra vertices at odd distance, and $t$ at even distance. The
corresponding generating functions depend on each other:
\begin{align}
  h_S(x,y) &= 1-\sum_q \Pi_q r_q + y \sum_q \Pi_q r_q \left[ h_T(x,y)
  \right]^{q-1},
  \label{eq:degcomponent_size_S} \\
  h_T(x,y) &= 1-\sum_q \Pi_q r_q + x \sum_q \Pi_q r_q \left[ h_S(x,y)
  \right]^{q-1},
  \label{eq:degcomponent_size_R}
\end{align}
and the function generating $\xi(s,t)$ is
\begin{equation}
  h_\xi(x,y) = x \sum_q \Pi_q \left[ h_S(x,y) \right]^q.
  \label{eq:degcomponent_size_P}
\end{equation}
This allows to compute $\xi(s,t)$ by taking partial derivatives in $x$
and $y$. As in the case of Eqs.~(\ref{eq:component_size_vertex_gf}) and
(\ref{eq:new_component_size_vertex_gf}), $h_\xi(x,y)$ is in general a
transcendental function and has to be solved numerically. However, in
some cases it can be solved analytically. In the case of 2-swap only
($\Pi_2=1$, $\Pi_{q\neq2}=0$), Eq.~(\ref{eq:degcomponent_size_P}) simplifies
to the closed form
\begin{equation}
  h_\xi(x,y) = \frac{x(1-r_1)^2(1+r_1y)^2}{(1-r_1^2xy)^2}.
\end{equation}
The probability $\xi(s,t)$ in Eq.~(\ref{eq:xi2}) is then
\begin{align}
  \xi(s,t)
  &= \frac{1}{s!t!} \left. \frac{\partial^s\partial^t h_\xi(x,y)}{\partial y^s
  \partial x^t} \right|_{x,y=0}
  \notag\\
  &= \binom{2}{1+s-t}(1-r_1)^2r_1^{s+t-1}t
  \label{eq:xi2_app}
\end{align}
if $|s-t|\le1$ and 0 otherwise.

Alternatively, for the case of a single target degree, $\eta_q^{(\rm rand)}$ can also be computed exactly up
to the $n$-th order in $r_{q-1}$ by the branching process depicted in
Figure~\ref{fig:clusters_eta23_net}. The process begins at step 0, with
$k_0=1$ vertices of degree $q$. At step 1, $k_1$ vertices out of $qk_0=q$ are
of degree $q$ with binomial probability
\[
\binom{q}{k_1} r_{q-1}^{k_1} (1-r_{q-1})^{q-k_1}.
\]
At following steps $i\ge2$ in the branching process, there are $(q-1)$ new
vertices for each previous vertex of degree $q$. Thus, in every step, $k_i$
vertices are of degree $q$ with probability
\[
\binom{(q-1)k_{i-1}}{k_i} r_{q-1}^{k_i} (1-r_{q-1})^{(q-1)k_{i-1}-k_i}.
\]
Operations are made on vertices at even steps. Note that every new step in the
branching process involves higher orders in $r_{q-1}$. Therefore, the
expansion of $\eta_{q}$ up to order $n$ is obtained by summing the
contributions of the first $n$ steps:
\begin{multline}
  \eta_q^{(\rm rand)}
  = \sum_{\{k_i\}} \frac{\sum_{i=0}^{\lfloor n/2\rfloor}k_{2i}}{\sum_{i=0}^nk_i}
  \binom{q}{k_1} \prod_{i=2}^n\binom{(q-1)k_{i-1}}{k_i}
  \\
  \times
  r_{q-1}^{\sum_{i=1}^nk_i} (1-r_{q-1})^{q+(q-2)(\sum_{i=1}^{n-1}k_i)-k_n}
  ,
  \label{eq:etaq_rand_N}
\end{multline}
where the sum in $\{k_i\}$ sums for $k_0=1$, $k_1=0,1,\dots,qk_0$ and
$k_{i\ge2}=0,1,\dots,(q-1)k_{i-1}$.

\bibliography{02}

\end{document}